\documentclass[structabstract]{aa}  % printer format
\usepackage{natbib}
\bibpunct{(}{)}{;}{a}{}{,} % to follow the A&A style
\usepackage{graphicx}
\usepackage{txfonts}
\newcommand{\exo}{EXO~2030+375}
\newcommand{\integral}{{\sl INTEGRAL}}
\newcommand{\ibis}{IBIS/ISGRI}
\newcommand{\jemx}{JEM-X}
\newcommand{\rxte}{{\sl RXTE}}
\newcommand{\exosat}{{\sl EXOSAT}}
\newcommand{\swift}{{\sl Swift}}

\begin{document}
   \title{Analyzing X-Ray Pulsar Profiles:}

   \subtitle{Geometry and Beam Pattern of EXO\,2030+375}

   \titlerunning{Pulse Profiles of EXO\,2030+375}

   \author{Manami Sasaki
          \inst{1}
          \and
          Dmitry Klochkov\inst{1}
          \and 
          Ute Kraus\inst{2}
          \and
          Isabel Caballero\inst{3}
          \and
          Andrea Santangelo\inst{1} 
          %\fnmsep\thanks{}
          }

   \institute{Institut f\"ur Astronomie und Astrophysik, 
              Universit\"at T\"ubingen,
              Sand 1, 
              72076 T\"ubingen, Germany,
              \email{sasaki@astro.uni-tuebingen.de}
         \and
              Institut f\"ur Physik und Technik,
              Universit\"at Hildesheim,
              Marienburger Platz 22,
              31141 Hildesheim, Germany
         \and
              CEA Saclay, DSM/IRFU/SAp -- UMR AIM (7158) 
              CNRS/CEA/Universit\'{e} P.Diderot --F-91191 Gif sur Yvette France
             %\thanks{}
             }

   \date{Received \today; accepted }

% \abstract{}{}{}{}{} 
% 5 {} token are mandatory
 
  \abstract
   {The pulse profiles of the transient Be/X-ray binary \exo\ show strong 
dependence on energy, as well as on its luminosity state, and are asymmetric in 
shape.}
   {We want to identify the emission components of the two magnetic poles in 
the pulsed emission to understand the geometry of the neutron star and its beam 
pattern.}
   {We utilize a pulse-profile decomposition method that enables us to find two 
symmetric pulse profiles from the magnetic poles of the neutron star. The
symmetry characteristics of these single-pole pulse profiles give
information about the position of the magnetic poles of the neutron star
relative to its rotation axis.}
   {We find 
a possible geometry for the neutron star in \exo\ through
%two solutions for the decomposition of the pulse profiles
%of \exo. Although the two solutions indicate two different possible geometries
%for the neutron star in \exo, the emission components that are responsible for 
%the main features in the pulse profiles are similar for the two solutions. 
%The 
the decomposition of the pulse profiles, which
suggests that one pole gets closer to the line of sight than the other and
that, during the revolution of the neutron star, both poles disappear behind 
the horizon for a short period of time. 
A considerable fraction of the emission arises from a halo while the pole is 
facing the observer and from the accretion stream of the other pole
while it is behind the neutron star, but the gravitational line bending makes
the emission visible to us.
}
   {}


   \maketitle
%
%________________________________________________________________

\section{Introduction}

\object{EXO 2030+375} is an accreting X-ray pulsar with a pulsation period of 
$\sim$42~s, which was discovered with \exosat\ in 1985 during a giant outburst 
\citep{1989ApJ...338..359P}.
A B0 Ve star was found as its counterpart in follow-up observations in the 
optical and infrared bands
\citep{1988A&A...202...81J,1987A&A...182L..55M,1988MNRAS.232..865C}.
During the giant outburst, \exo\ showed a spin-up of $-P/\dot{P} \approx 30$~yr 
\citep{1989ApJ...338..359P} and quasi-periodic oscillations with a frequency 
of $\sim$0.2~Hz \citep{1989ApJ...346..906A} interpreted as caused by
the formation of an accretion disk. Detailed analyses have shown that 
its rate of pulse-period change $\dot{P}$, energy spectrum, and pulse profile 
are strongly luminosity dependent 
\citep{1989ApJ...338..373P,1989ApJ...338..359P,1993ApJ...414..302R}.
The orbital period is 46 days \citep{2002ApJ...570..287W}, and a normal outburst 
has been detected for nearly every periastron passage since 1991 
\citep{2005ApJ...620L..99W}.
In 2006, \exo\ underwent the first giant outburst since its discovery in 1985 
\citep{2006ATel..843....1C,2006ATel..861....1K,2006ATel..868....1M}, 
during which it reached a maximum luminosity of $L_{1 - 20~{\rm keV}} \approx 
1.2 \times 10^{38}$~erg~s$^{-1}$ \citep{2008A&A...491..833K}
and again showed a strong spin-up.
Rossi X-ray Timing Explorer (\rxte) monitored \exo\ extensively during the
2006 giant outburst \citep{2008ApJ...678.1263W}. The source was also observed
by the INTErnational Gamma Ray Astrophysics Laboratory 
\citep[\integral,][]{2003A&A...411L...1W} and \swift\ 
\citep{2004ApJ...611.1005G}. The spectra indicate a cyclotron 
absorption line \citep{2007A&A...464L..45K,2008ApJ...678.1263W}.
\citet{2008A&A...491..833K} have shown that the spectrum of \exo\ changes
with pulse phase, suggesting a fan beam geometry during the maximum,
while towards the end of the giant outburst, it changes to a 
combination of a fan beam and a pencil beam. 

In X-ray pulsars, a neutron star accretes matter from a companion star
via stellar wind or Roche lobe overflow. The accreted matter is channeled
along the field lines of the strong magnetic field of the neutron star onto
the magnetic poles. X-ray emission from the neutron star is produced in 
regions around the two magnetic poles. As the magnetic dipole axis is most
likely inclined against the rotation axis of the neutron star, a distant
observer sees pulsed emission.
X-ray pulsars exhibit a wide variety of pulse shapes that differ from 
source to source. Generally, high-energy pulses have simpler shapes than 
low-energy pulses 
\citep[][and references therein]{1983ApJ...270..711W,1989ESASP.296...57F,1997ApJS..113..367B}.
If one assumes an axially symmetric geometry for the two emission regions of 
the neutron star in a dipole configuration, the observed pulse profile 
should be symmetric. However, the observed pulse profiles typically show an
asymmetry. 
To explain the asymmetric shape of the total pulse profile, a distorted 
magnetic dipole field in which the two magnetic poles are not located opposite
each other have been discussed \citep{1989ApJ...338..373P,1991MNRAS.251..203L,1993ApJ...406..185R,1995ApJ...444..405B}. 
\citet{1995ApJ...450..763K} shows that, starting from the observed, 
asymmetric pulse profile, it is possible to disentangle the 
contribution of the two emission regions of the neutron star. 
Once the pulsed emission from each of the poles has been obtained, one can derive
the geometry of the neutron star. This again allows us to 
construct the beam pattern, i.e., the flux distribution from one emission 
region. Using this pulse-profile decomposition method, 
\citet{1996ApJ...467..794K} have analyzed the pulse profiles of Cen~X-3 and 
find indications of both pencil and fan beam.
In the case of Her~X-1, the results of the pulse-profile decomposition by 
\citet{2000ApJ...529..968B} have not only shed light on the beam pattern of 
the magnetic poles, but have also confirmed that a warped and tilted accretion 
disk attenuates the emission from one pole of the neutron star. 
For A~0535+26, the reconstructed beam pattern suggests that the
emission comes from a hollow column plus a halo of scattered radiation
on the neutron star surface \citep{Caballero2010}.

In this paper we present the analysis of the energy-resolved pulse profiles
of \exo\ utilizing the decomposition method developed by 
\citet{1995ApJ...450..763K}. Section \ref{data} gives an overview of the
data used for our analysis and Sect.\,\ref{analysis} describes the analysis
and the results obtained with the pulse-profile decomposition method. 
The results are discussed in Sect.\,\ref{discussion}. Section \ref{summary} 
summarizes the possible geometry of the neutron star and the origin
of the observed emission.

\section{Data}\label{data}

%--------------------------------------------------------- folded lightcurves
\begin{figure*}
\centering
\hspace{5mm}
%{\scriptsize $L_{1 - 20~{\rm keV}} = 1 \times 10^{38}$~erg~s$^{-1}$}\\[-5mm]
%\includegraphics[width=.45\textwidth,angle=270,clip=]{pulseprofiles_high_bw_1.eps}
%\hspace{-1mm}
%\includegraphics[width=.45\textwidth,angle=270,clip=]{pulseprofiles_high_bw_2.eps}\\[3mm]
%\hspace{5mm}
%{\scriptsize $L_{1 - 20~{\rm keV}} = 7 \times 10^{37}$~erg~s$^{-1}$
%\hspace{15mm}
%$L_{1 - 20~{\rm keV}} = 6 \times 10^{37}$~erg~s$^{-1}$ 
%\hspace{15mm}
%$L_{1 - 20~{\rm keV}} = 1 \times 10^{37}$~erg~s$^{-1}$}\\[-5mm]
\includegraphics[width=.45\textwidth,angle=270,clip=]{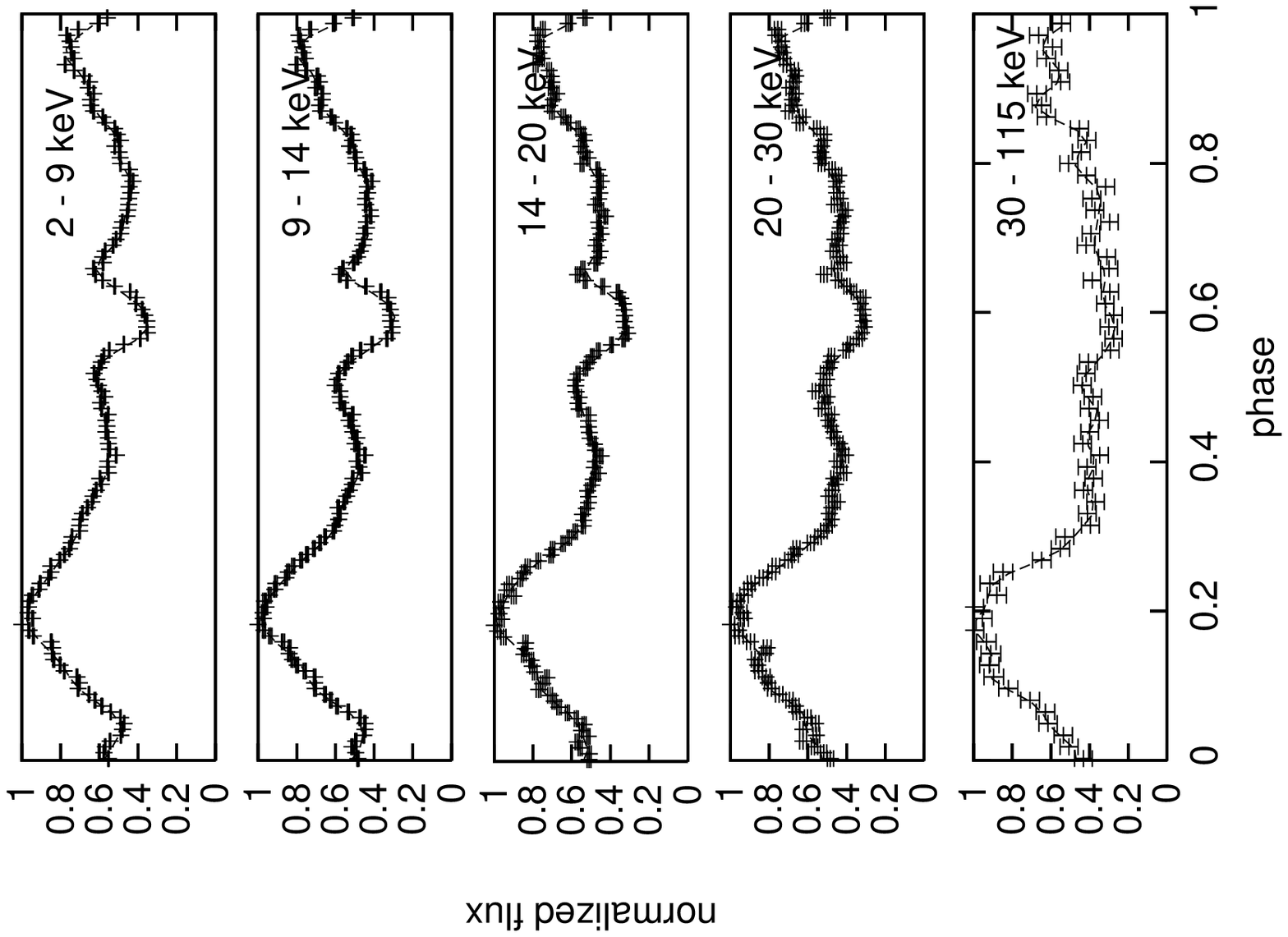}
%\hspace{1mm}
\includegraphics[width=.45\textwidth,angle=270,clip=]{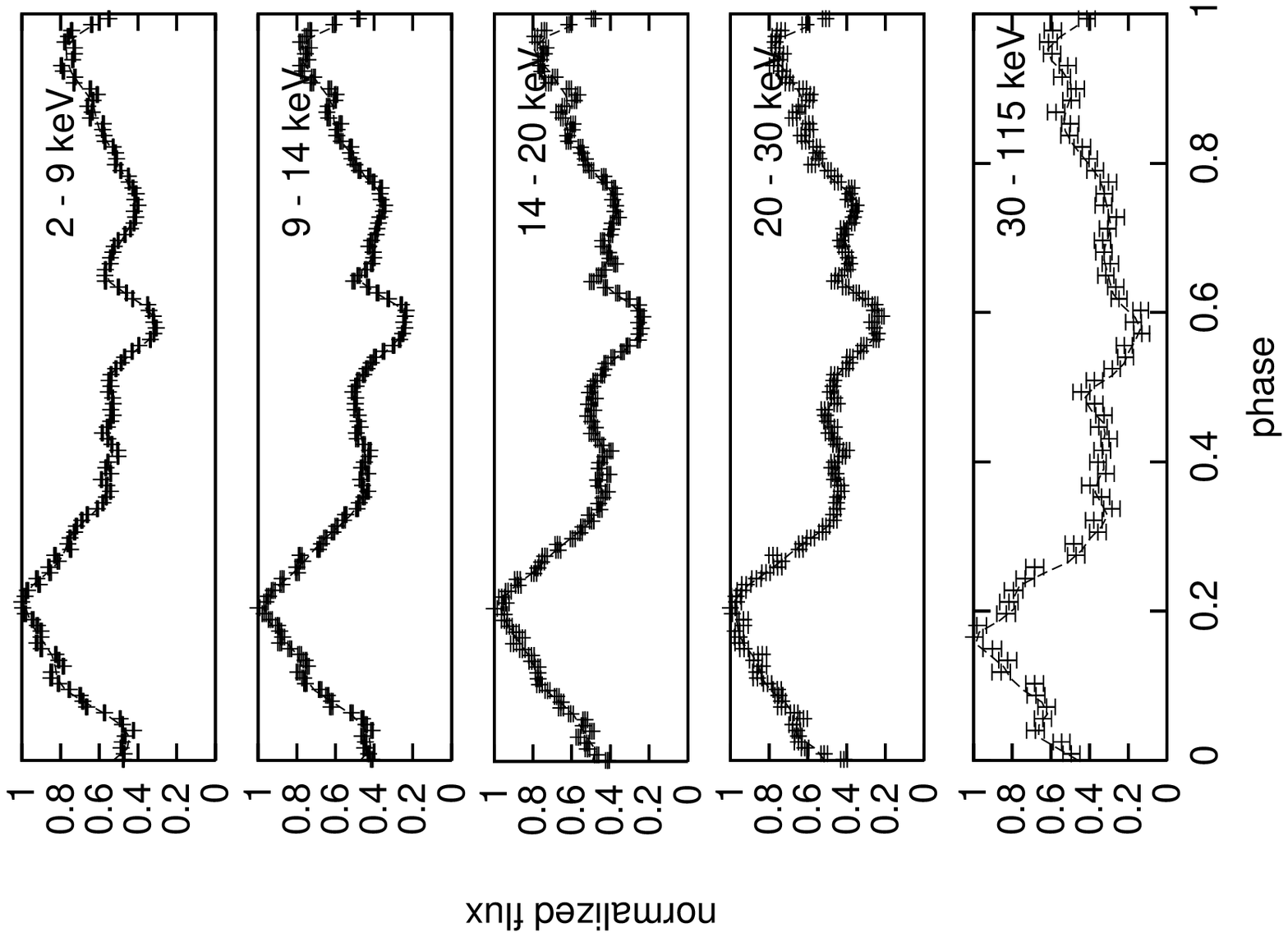}
%\hspace{1mm}
%\includegraphics[width=.37\textwidth,angle=270,clip=]{pulseprofiles_low_bw.eps}
% %%BoundingBox: 138 50 554 415
\caption{
%        {\it Upper panel:}
%        Pulse profiles of \exo\ during the maximum of the
%        giant outburst in 2006 observed with \integral.
%        {\it Lower panel:}
        Pulse profiles of \exo\ during the rise (left) and the decay (right)
%        middle) 
        of the giant outburst in 2006 observed with \rxte.
%        , as well as the 
%        pulse profiles at the end end of the decay of the 
%        giant outburst in 2006 observed with \integral\ (right).
%        The luminosity of the source is given at the top of the panels.
        }
\label{folded}
\end{figure*}
%______________________________________________________________

\exo\ experienced a giant outburst in 2006, during which the source was
monitored continuously by \rxte\ and was also observed by \integral.
We have used the pulse profiles obtained with the Joint European X-Ray Monitor 
\citep[\jemx,][]{2003A&A...411L.231L} and the imaging system \ibis\ 
\citep{2003A&A...411L.131U} as presented in Figs.\,2 and 8
of \citet{2008A&A...491..833K}.

\subsection{\rxte\ observations}

For better statistics, we also used publicly available archival data
from two observations with \rxte\ during the rise and the decay of the giant
outburst when \exo\ showed about half of the maximum luminosity.
The observations took place on June 28 and September 17, 2006
(observation IDs 91089-01-07-00 and 91089-01-19-01) when the luminosity of the 
source was $L_{1 - 20~{\rm keV}} = 7 \times 10^{37}$~erg~s$^{-1}$ and 
$6 \times 10^{37}$~erg~s$^{-1}$, respectively.
We started from the event files obtained with the Proportional Counter Array 
\citep[PCA,][]{1996SPIE.2808...59J}. 
We used the event encoded mode files in oder to have optimum binning. After 
filtering good time intervals and applying bitmasks, we created lightcurves with
a time binning of 0.125~s in the following spectral bands: 2 -- 9~keV, 
9 -- 14~keV, 14 -- 20~keV, 20 -- 30~keV, and 30 -- 115~keV. After background
subtraction, the lightcurves were corrected to solar barycenter and for orbitary 
motion of the binary. 
After folding the lightcurves with periods measured for each observation, 
we obtained pulse profiles with 128 phase bins. Phase 0.0 was fixed to agree 
with the \integral\ pulse profiles of \citet{2008A&A...491..833K}.
%The location of 
%phase 0.0 in the folded lightcurves is defined in such a way that it is the 
%same as in the \integral\ data: we define the minimum before the pronounced 
%maximum as phase 0.0 and determine its position in the lightcurves by fitting
%a Gaussian. The phase definition is also cross-checked using the sharp 
%minimum at about phase 0.6.
The folded lightcurves from the \rxte\ observations are shown in Fig.\,\ref{folded}.
% (lower panel).

During the two \rxte\ observations shown here, the luminosity of \exo\
was comparable. As can be seen in Fig.\,\ref{folded}, the pulse profiles 
before the maximum and after the maximum of the giant outburst are very 
similar, corroborating
that the shape of the pulse profiles only depends on the luminosity state
\citep{1989ApJ...338..373P}.

\section{The analysis}\label{analysis}

\subsection{The method}\label{meth}

A detailed description of the pulse-profile decomposition method, which is 
based on a backward tracing of the emission, can be found in 
\citet{1995ApJ...450..763K}.
All major steps of our analysis and the criteria applied to obtain the best 
solution are described in the Appendix.

The basic assumption of the method is that
%, to explain the asymmetric shape of the pulse profiles, 
the magnetic dipole field of the neutron star is 
distorted in such a way that the two magnetic poles do not lie on a straight 
line through the center of the neutron star.
Therefore, even though the emission from each pole is axisymmetric, the sum
of the emission from both poles results in an asymmetric pulse profile.
Using Fourier analysis, we model the observed asymmetric pulse profiles with 
two symmetric functions 
$f_{1,2}$ to search for symmetry points $\Phi_{1,2}$ in the pulse profiles 
and their offset $\Delta$ (see Appendix for details). For each observation 
and energy range, the functions $f_{1}$ and $f_{2}$ correspond to the two 
single-pole pulse profiles that in total add up to the observed asymmetric 
pulse profile.
Each symmetry point corresponds to the pulse phase where the respective pole
is either closest to or most distant from the observer's line of sight.
From the two symmetry
points and functions, we then derive the location of the emission regions and
the beam pattern.

\subsection{Decompositions}

%______________________________________________ single-pole pulse profiles
\begin{figure*}
\centering
\hspace{8mm}{\footnotesize\sf Solution 1}\\[-1mm]
\includegraphics[width=0.7\textwidth,angle=0,clip=]{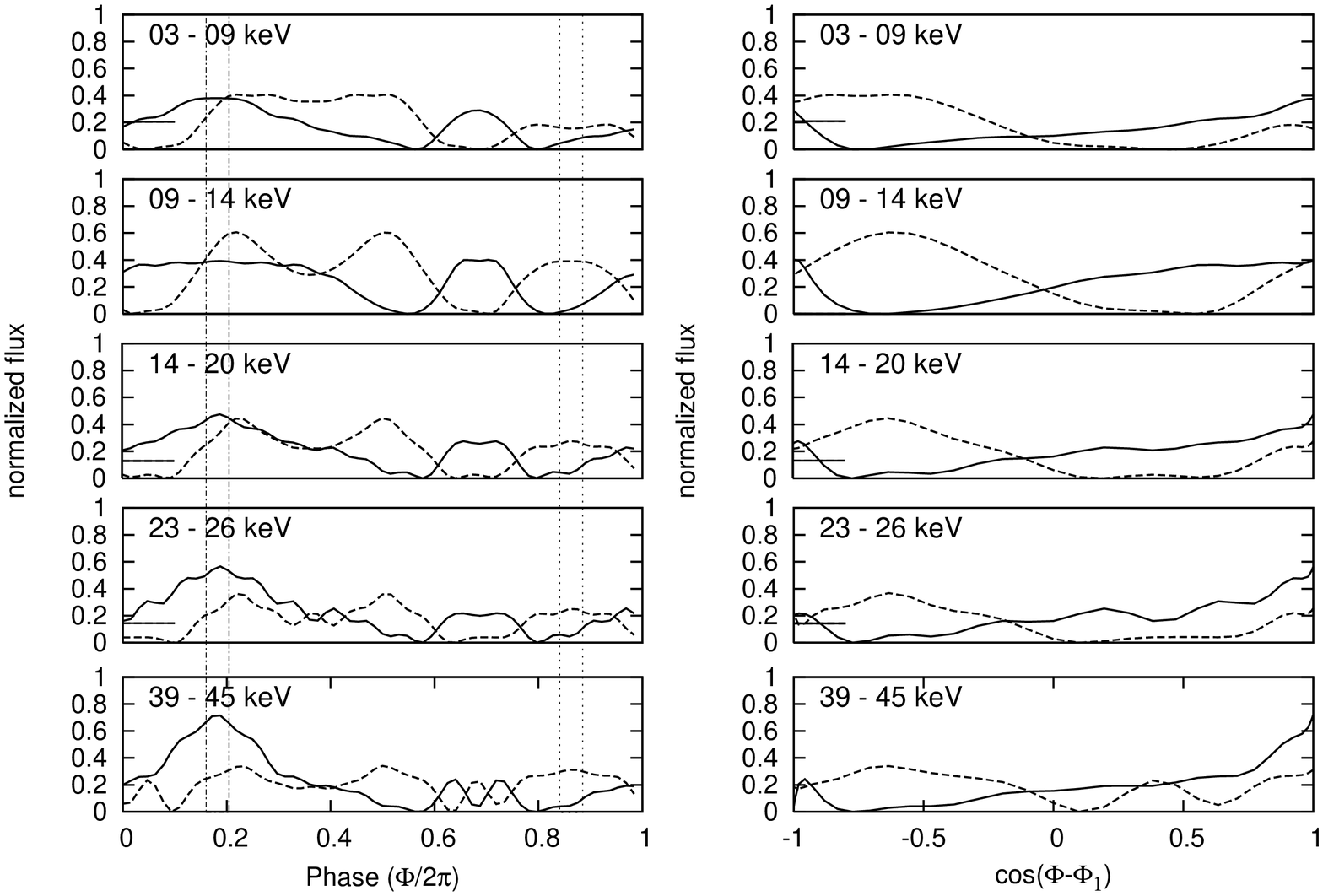}
%\\[-3mm]
%\hspace{8mm}{\footnotesize\sf Solution 2}\\[-1mm]
%\includegraphics[width=0.7\textwidth,angle=0,clip=]{sppbp_A2.eps}
\caption{
A selection of single-pole pulse profiles (left) and corresponding 
beam patterns (right) obtained by the decomposition 
method for the $\Phi_{1}$-$\Delta$ set for solution 1 
%(upper panel) and solution 2 (lower panel) 
for the energy bands 3 -- 9~keV, 9 -- 14~keV, 
14 -- 20~keV, 23 -- 26~keV, and 39 -- 45~keV of the \integral\ observation near 
maximum. The solid line is used for emission from the first pole, dashed line
for the second pole. The vertical dash-dotted lines indicate the ranges
for the symmetry point $\Phi_{1}$, while the dotted lines show the ranges for 
$\Phi_{2}$ obtained from pulse profiles at different energies and observations.
}
\label{sppbp}
\end{figure*}
%______________________________________________________________

We have a total of 26 pulse profiles from \integral\ and \rxte\ observations. 
To perform pulse-profile decomposition, the maximum in each of the 26 
pulse profiles is normalized to unity. Thereafter each pulse
profile is modeled with two symmetric functions $f_{1}$ and $f_{2}$ based on
Fourier analysis. After applying criteria 1 ({\it positive flux}, see 
Appendix \ref{meth_decomp}) and 2 ({\it no ripples}),
we obtain a large number of possible values for the parameters $\Phi_{1}$
and $\Delta$. However, after combining the results for all pulse profiles
(criterion 3), only two interesting solution regions remain in the parameter 
space of $\Phi_{1}$-$\Delta$. We call these solutions 1 and 2 and perform 
further analysis with these two possible solutions.

For each observation and energy band, the functions $f_{1}$ and $f_{2}$ 
correspond to the two single-pole pulse profiles that in total add up to the 
observed asymmetric pulse profile. The Fourier analysis finds more than
one possible set of $\Phi_{1}$ and $\Delta$ within a small region for one
solution. For each total pulse profile, we have to look at the
different sets of the single-pole pulse profiles to decide which one is 
consistent with the single-pole pulse profiles at other energies.
The pulse profiles of the different observations are studied separately, 
because one should see a correlation between the
different energy bands of one luminosity state, but not necessarily between
two different observations. Figure \ref{sppbp} shows some of the selected 
single-pole pulse profiles and the derived beam patterns for each pole
for solution 1. Within one observation, one can see an energy-dependent 
evolution of the single-pole pulse profiles. 
The values of $\Phi_{1}$ and $\Delta$ for solutions 1 and 2 are 
(65\degr\ $< \Phi_{1}$ $<$ 75\degr, 63\degr\ $< \Delta$ $<$ 70\degr) and 
(70\degr\ $< \Phi_{1}$ $<$ 80\degr, 81\degr\ $< \Delta$ $<$ 88\degr), 
respectively. 
The parameters for the two solutions are listed in Table \ref{tabpa}.

%_____________________________________________________________
%
\begin{table}
\begin{minipage}[t]{\columnwidth}
\caption{Parameters obtained from the decomposition for solutions 1 and 2}
\label{tabpa}
\begin{center}
\begin{tabular}{lccccccc}
\hline\hline\noalign{\smallskip}
 & $\Phi_{1}$ & $\Delta$ & $a$ & $b$ & $\Theta_{1}\,^1$ & $\Theta_{2}\,^1$ & $\delta\,^1$ \\
\hline\noalign{\smallskip}
1 & 65\degr\ -- 75\degr\ & 63\degr\ -- 70\degr\ & --2.1 & 1.0 & 39\degr\ & 141\degr\ &40\degr\ \\
%aus E0309.wi_209_100 (A1)
2 & 70\degr\ -- 80\degr\ & 81\degr\ -- 88\degr\ & --0.1 & 1.0 & 87\degr\ & 93\degr\ & 85\degr\ \\
%aus E0309.wi_100_100 (A2)
\hline                    
\end{tabular}
\end{center}
\vspace{-2mm}
$^1$ Assuming $\Theta_{0}$ = 50\degr.
\end{minipage}
\end{table}
%_____________________________________________________________

\subsection{Overlaying beam patterns}\label{asymbp}

As the neutron star rotates, the angle between the axis through one magnetic
pole and the line of sight $\theta$ changes with phase, i.e., with 
rotation angle $\Phi$.
The decomposition has provided us with beam patterns as seen by the distant 
observer for each emission region as functions of the phase.
Now we compare the two beam patterns derived from the two single-pole pulse
profiles of each observed pulse profile and search for a range
in $\cos(\Phi-\Phi_{1})$, and thus $\theta$, in which the two beam patterns 
seem to show the same emission (see Appendix \ref{meth_bp}). 
We try to overlay the two beam patterns by using the relation:
\begin{equation}\label{ab}
\cos(\Phi-\Phi_{1}) = 
a + b~\cos(\tilde{\Phi}-\Phi_{2}), \hspace{3mm} b > 0.
\end{equation}
For solution 1 we are not able to find an overlap of the single-pole beam 
patterns. It is more likely that, in this case, the geometry only allows the
observer to see two different parts of the total beam pattern. To assemble the
two parts of the beam pattern, we use $a = -2.1$ and $b = 1.0$ 
(see Fig.\,\ref{bpobs}, upper panel).
For solution 2, we find an overlap of the single-pole beam patterns for
$a = -0.1, b = 1.0$, although the beam patterns do not seem to match perfectly.

%_____________________________________________ asymptotic beam patterns, + (angle)
\begin{figure*}
\centering
\hspace{11mm}{\footnotesize\sf Solution 1}\\[-2mm]
\includegraphics[width=0.5\textwidth,angle=270,clip=]{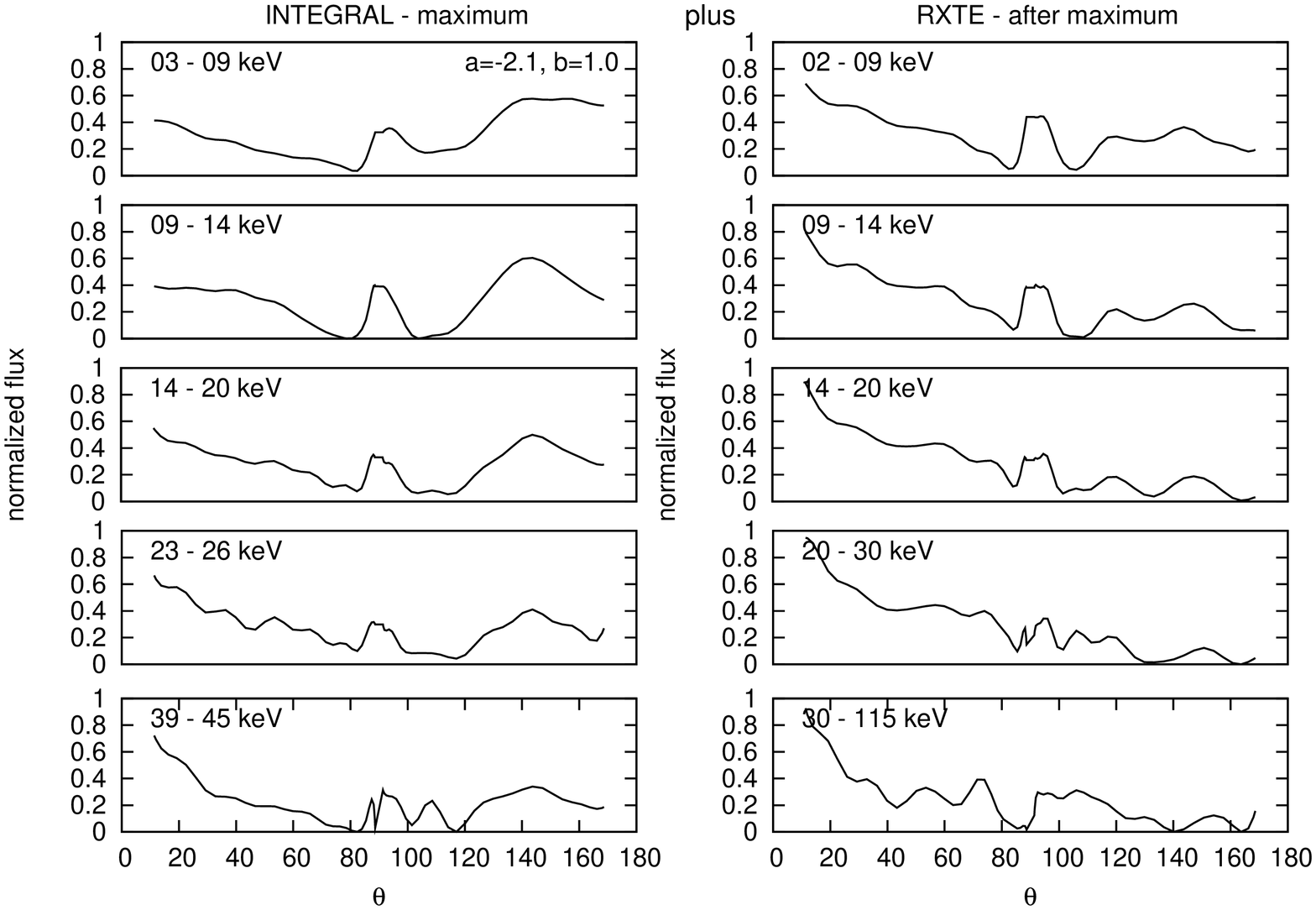}\\
\includegraphics[width=0.5\textwidth,angle=270,clip=]{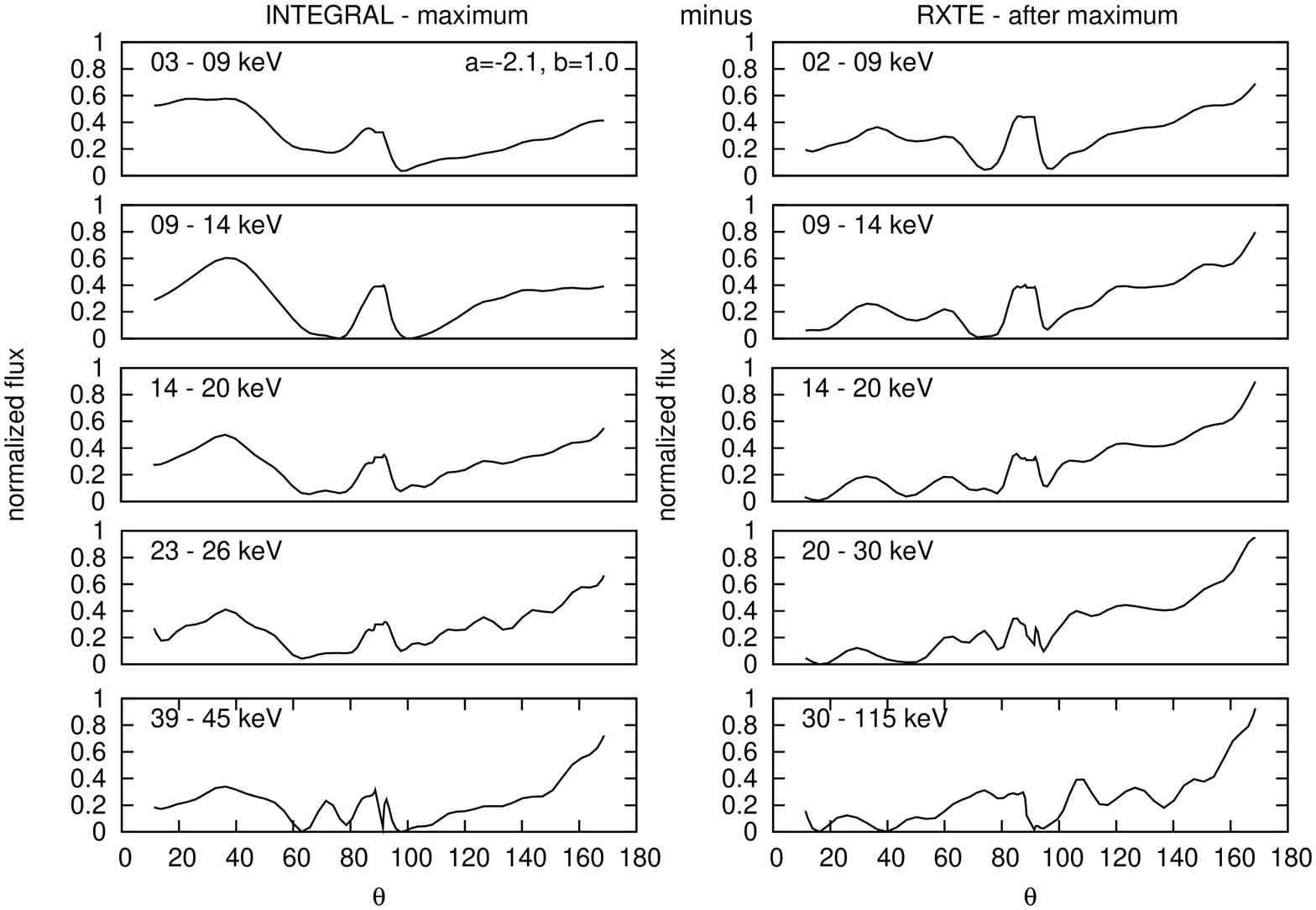}
\caption{Asymptotic beam patterns of the {\tt plus} and {\tt minus} solutions 
of solution 1 seen by the distant observer plotted over the angle $\theta$ 
between the first magnetic pole and the line of sight. 
The left panel shows the beam patterns of 
solution 1 of the observation near maximum as shown in Figure \ref{sppbp}.
The right panel shows a selection of additional data from the observation 
after maximum with similar energy bands. 
%For the definition of the {\tt plus} and {\tt minus} solutions 
%see Sect.\,\ref{meth_bp}.
}
\label{bptheta1}
\end{figure*}
%______________________________________________________________

%%_____________________________________________ asymptotic beam patterns (angle)
%\begin{figure*}
%\centering
%\hspace{11mm}{\footnotesize\sf Solution 2}\\[-2mm]
%\includegraphics[width=0.5\textwidth,angle=270,clip=]{bp.asymp.A2.eps_pages1}\\[0mm]
%\includegraphics[width=0.5\textwidth,angle=270,clip=]{bp.asymp.A2.eps_pages2}
%\caption{Same as in Figure \ref{bptheta1}, but for the solution 2.
%For this solution, the beam patterns in the overlap range are averaged.}
%\label{bptheta2}
%\end{figure*}
%%______________________________________________________________

A total beam pattern can be reconstructed from the
beam patterns calculated from the single-pole pulse profiles.
As shown in \citet{1995ApJ...450..763K}, there is an ambiguity in the
relation between $\Phi$ and $\theta$,
as each single-pole pulse profile has two symmetry points
at $\Phi_{i}$ and $\Phi_{i} + \pi$. Therefore, the solutions cannot tell us
which ends of the sections of the beam patterns belong to, e.g., the lower
values of $\theta$. For each
set of $\Phi_{1}$ and $\Delta$ we obtain two possible solutions (called 
{\tt plus} and {\tt minus}) for the total beam pattern. The decomposition 
method cannot tell us which one is the real solution. We have to take results 
from other measurements of the source into consideration, e.g., luminosities 
or spectra that will give insight into the emission processes, to decide which 
of the two is the real solution.  
The reconstructed asymptotic beam patterns for solution 1 as seen by
the distant observer are shown in Fig.\,\ref{bptheta1}.
% and \ref{bptheta2}.
Here the beam patterns are plotted over the angle $\theta$ between the normal
at the first magnetic pole and the line of sight.

%\subsubsection{Not Matching Beam Patterns for Solution 2}

That the beam patterns of the single-pole pulse profiles do not match 
in the overlap region of solution 2 %(see Fig.\,\ref{bpobs}) 
might indicate that the emission from the two magnetic poles are not identical 
and cannot be described with one local beam pattern. Therefore, we take the mean 
of the two beam patterns and model a total averaged beam pattern, which we use 
to reconstruct the visible total pulse profile, i.e., assuming equal local 
emission pattern for the two poles. 
%The resulting pulse profiles are shown with dashed lines 
%in Figure \ref{totalppA2}. As one can see, the modeled
%pulse profile well reproduces the main features like the major peaks at 
%phases 0.2 and 0.95 as well as the minimum at phase 0.6. 
%However, there are significant deviations from the real total
%pulse profiles. 
The reconstructed pulse profiles (Fig.\,\ref{totalppA2}) show significant 
deviations from the observed profiles.
%A possible explanation is that the local emission patterns of the two magnetic 
%poles are not identical, causing the small-scale fluctuations in the pulse 
%profile.
%
%As will be shown in Section \ref{geomet}, one can derive the geometry for the
%neutron star from the results of the decomposition analysis.
%The geometry that we obtain for solution 2 is
%very intriguing: for almost all possible $\Theta_{0}$, all the angles
%$\Theta_{1}, \Theta_{2}$, and $\delta$ are $\sim$90\degr (see Table \ref{tabpa}).
%% (see Fig.\,\ref{geometry}, right panel). 
%This means that the two magnetic poles are located near the equator of
%the neutron star and have an angle of $\sim$90\degr\ between each other.
%This is a very extreme geometry, completely different from the classical
%antipodal configuration. 
%In principle, we cannot rule such a strongly distorted geometry out especially 
%not for young neutron stars like those expected in Be/X-ray binary systems, as 
%they might have experienced some anisotropic conditions while their birth.
%
In addition, this solution yields a very extreme 
geometry with the two magnetic poles located near the equator of the neutron star,
forming an angle of $\sim$90\degr\ between each other (see Table \ref{tabpa}).
Although, in principle, we cannot rule such a strongly distorted geometry out,
especially not for young neutron stars like those expected in Be/X-ray binary 
systems, 
as they might have experienced some anisotropic conditions while their birth,
this rather unlikely geometry also suggests that solution 2
is not appropriate for \exo. Therefore, in the following, we focus on 
the discussion of solution 1.
%The visible angle in this case are $\sim$30\degr\ -- 95\degr\ for 
%$\Theta_{0}$ = 50\degr, i.e., narrower than for solution 1. If one keeps
%this in mind, the beam patterns of solution 2 as shown in the polar diagrams
%of Figure \ref{bpintrA2} are not significantly different from the
%beam patterns of the {\tt plus} solution of solution 1 
%in Figure \ref{bpintrA1}. 

\subsection{Geometry of the neutron star}\label{geomet}

%______________________________________________ geometry of the neutron star 
\begin{figure}
\centering
\includegraphics[width=0.35\textwidth,angle=90,clip=]{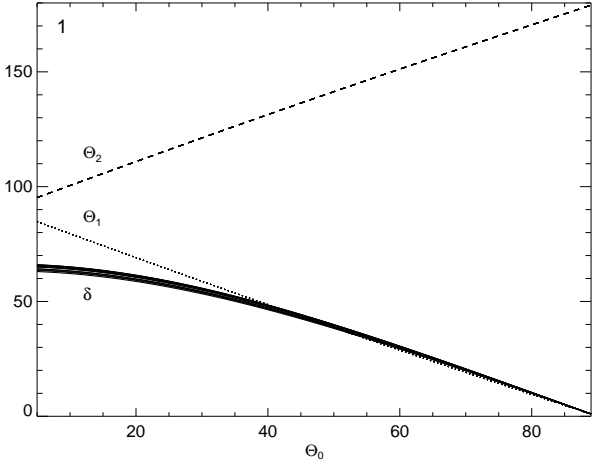}
\caption{Relation between the observing angle $\Theta_{0}$
and the angles $\Theta_{1}$ (dotted line), $\Theta_{2}$ (dashed line), and 
$\delta$ (range of possible values) for solution 1.}
% (left) and solution 2 (right).}
\label{geometry}
\end{figure}
%______________________________________________________________

To derive the exact geometry of the neutron 
star, i.e., to determine the polar angles of the magnetic poles $\Theta_{1}$
and $\Theta_{2}$ as well as the offset angle $\delta$, we need to know the
inclination angle of the rotation axis of the neutron star. However, this 
angle is not known for most of the neutron stars, in particular not for
\exo. Figure \ref{geometry} shows the dependence of the angles $\Theta_{1}$,
$\Theta_{2}$, and $\delta$ on the angle $\Theta_{0}$ between the rotation axis
of the neutron star and the line of sight of the observer. 
To convert the phase parameter $\cos(\Phi-\Phi_{1})$ into the angle $\theta$
between the first magnetic pole and the observer, we have to assume an 
inclination angle $\Theta_{0}$ with respect to the rotation axis. 
Here, we use $\Theta_{0}$ = 50\degr, corresponding to $\Theta_{1}$ = 39\degr, 
$\Theta_{2}$ = 141\degr (see Fig.\,\ref{geometry} and Table \ref{tabpa}).

\section{Discussion}\label{discussion}

%The decomposition of the total pulse profiles obtained from four observations
%with \integral\ and \rxte\ during the giant outburst of \exo\ in 2006 has 
%provided us with two possible solutions for the symmetry points in the pulse 
%profiles $\Phi_{1}$ and $\Phi_{2}$ and beam patterns corresponding to the 
%single-pole pulse profiles.
%As discussed in the last section, we consider solution 1 as the most 
%likely result for \exo.
%Here, 
In this section we want to further examine the results of the decomposition 
method and present the possible geometry of the neutron star.

\subsection{Disentangling the emission components}

%____________________________ components in the single-pole pulse profiles 
\begin{figure}
\centering
%\includegraphics[width=0.49\textwidth,clip=]{onespp.A1.comp.eps}
%%\includegraphics[width=0.49\textwidth,clip=]{onespp.A2.comp.eps}
%\caption{Decomposed single-pole pulse profiles of the data taken near the
%maximum of the giant outburst for the lowest energy band (same as the upper 
%left diagram of the upper [solution 1]
%%, here left] and lower [solution 2, here right]
%panels in Fig.\,\ref{sppbp}) with `pencil beam' (mostly forward directed) emission 
%component and `fan beam' (wide, directed mostly to the side) emission 
%component from each pole. 
%`Fan beam' emission is hatched with vertical stripes for the first pole, 
%whereas the diagonally striped component shows the `fan beam' from the second 
%pole. 
%%In the case of solution 1, 
%The second pole never faces towards
%the observer, therefore, the observer only see the `fan beam' emission. 
%
\includegraphics[width=0.49\textwidth,clip=]{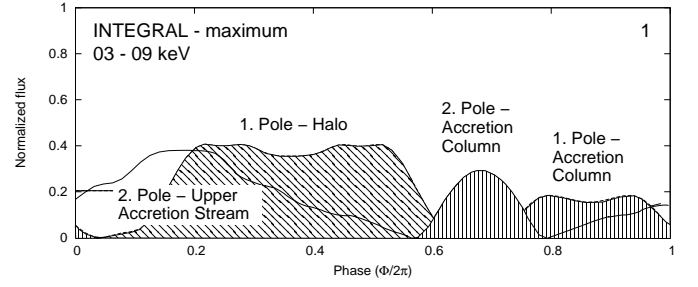}
\caption{
Decomposed single-pole pulse profiles of the data taken near the
maximum of the giant outburst for the lowest energy band (same as the upper 
left diagram of the upper [solution 1] panels in Fig.\,\ref{sppbp}) with 
proposed emission components derived from the beam patterns of the {\tt minus}
solution (Fig.\,\ref{bptheta1}, lower panel). 
We suggest that the emission from the first pole consists of emission from the 
halo (diagonally striped component) and from the accretion column (vertically 
striped). The dominant emission from the
second pole seems to have its origin in the scattered photons in the upper 
accretion stream, which is visible while the second pole is behind the horizon
of the neutron star due to gravitational light bending, 
while there is also a contribution from the accretion column
(vertically striped). In this solution, the first pole gets 
closer to the line of sight than the second.
}
\label{sppcomp}
\end{figure}
%______________________________________________________________

For solution 1, there is no overlap between the two beam patterns obtained
from the single-pole pulse profiles as shown in Sect.\,\ref{asymbp}. 
It means that only a part of the emission is seen from 
each pole during the revolution of the neutron star. 
By putting the two visible parts 
together, we obtain the total beam pattern of the emission around one magnetic
pole. In doing so, we make the assumption that the two magnetic poles have
the same emission pattern. 
The {\tt plus} solution (Fig.\,\ref{bptheta1}, upper panel) can be 
described as a composition of a forward directed emission (towards 0\degr) 
that is more pronounced at higher energies and an extended, relatively soft 
emission ($\sim$80\degr\ -- 180\degr). The harder, forward-directed emission 
indicates a pencil beam. The relative flux of the softer emission component 
is higher in the data from the observation performed near the maximum 
compared to the one during the decay and can be 
interpreted as a fan beam, in agreement with the largely accepted picture 
that an optically thick accretion column is formed during the giant outburst 
\citep[][and references therein]{1983ApJ...270..711W}.
The {\tt minus} solution has a soft emission at $\theta \la$ 60\degr and a 
harder emission at $\theta \ga$ 120\degr\ that increases for larger $\theta$. 
As newest calculations by Kraus et al.\ (in prep.) have shown, reprocessing of 
photons in the upper accretion stream creates a significant emission component 
that dominates the beam pattern at higher energies and can be observed while 
the pole, hence the accretion column, 
is on the other side of the neutron star (`anti-pencil'). In addition,
the emission from a halo that is formed by scattered photons at the bottom 
of the accretion column dominates the beam pattern at lower energies and 
at lower $\theta$ and is stronger in the data near the maximum of
the giant outburst.

Figure \ref{sppcomp} shows how the emission from the two poles contributes 
to the pulse profiles for the decompositions of the low energy band 
data taken at the maximum of the giant outburst in the case of the 
{\tt minus} solution.
Let us assume that the first magnetic pole with the polar angle $\Theta_{1}$
gets closer to the line of sight than the second magnetic pole, i.e.,
$\Theta_{0}-\Theta_{1} \le \Theta_{2}-\Theta_{0}$.
At phase 0.0, both poles are right behind the horizon of the neutron star:
the first pole is going to reappear, the second pole is turning farther away 
from the observer. 
At phase $\sim0.1$ the first pole becomes visible, and from then on the 
emission from the halo of the first pole makes the largest contribution until
the pole disappears behind the horizon at $\sim0.6$. When the first pole, 
which is closer to the observer's line of sight, is right at the horizon
of the neutron star and its accretion column is seen from the side, a minimum
is likely to be observed in the pulse profile (see also Sect.\,\ref{emorigin}).
At the major maximum at phase 0.15 -- 0.25, the second magnetic pole is
behind the horizon of the neutron star and the scattered and gravitationally
bent photons from the upper accretion stream cause the pronounced increase in 
flux. 
The second pole that is rotating on a circle farther away from 
the line of sight than the first pole comes back to the front side at about 
phase 0.45 and is closest to the observer at about phase 0.7. However,
since the line of sight is closer in latitude to the first pole than to the
second pole, the observer never gets as close to the surface normal of 
the second pole as to that of the first pole. The main emission seen from the
second pole while it is on this side of the horizon comes directly from the
accretion column.

\citet{1989ApJ...338..373P} have modeled the luminosity dependent
pulse profiles from the first observed giant outburst in 1985 by 
assuming a fan beam and a pencil beam component for the two magnetic poles
based on a model by \citet{1981A&A...102...97W}. They obtain a fit 
for all ten pulse profiles for different luminosities ranging from
0.1 -- 10.0 $\times 10^{37}$~erg~s$^{-1}$ with some residuals and find
that the fan beam mainly produces the major peak at phase 0.2, whereas
the peak at phase 0.95 can be ascribed to a pencil beam. The
way they have chosen phase 0.0 is different than in our work, resulting 
in a shift by about 0.4. The best-fit parameters of \citet{1989ApJ...338..373P}
correspond to $\Theta_{1}$ = 70\degr, $\Theta_{2}$ = 110\degr, and 
$\delta$ = 70\degr, assuming $\Theta_{0}$ = 25\degr. This result is consistent
with our solution 1 for the case of $\Theta_{0}$ = 25\degr\ (see 
Fig.\,\ref{geometry}).
%, left diagram).

\subsection{Comparison to phase-resolved spectral analysis}

\citet{2008A&A...491..833K} analyzed the same \integral\ data as used in 
this work and performed pulse phase resolved spectroscopy. For the data of the
maximum, they find that the spectrum is harder at the main peak (phase 0.2).
They conclude that the observer might be seeing Compton scattered photons from 
the optically thick accretion column. 
In the data taken at the end of the decay, a new peak appears at about phase
0.95. At this point, the spectrum is again harder, leading to the conclusion
that the line of sight of the observer is closest to the magnetic field lines.
This picture is in good agreement with the solution 1 of the decomposition, 
in which the 
emission of the upper accretion stream of the second pole, which
is visible owing to gravitational light bending, has its maximum at
phase $\sim$0.2 while the halo of the first pole also has substantial 
emission, thus contributing significantly to the main peak.
At phase $\sim$0.95 the scattered emission from
the upper accretion stream of the second pole starts to increase,
while the emission from the accretion column of the first
pole also contributes to the observed emission.

\subsection{Intrinsic beam patterns}

%______________________________________________ intrinsic beam patterns A1
\begin{figure*}
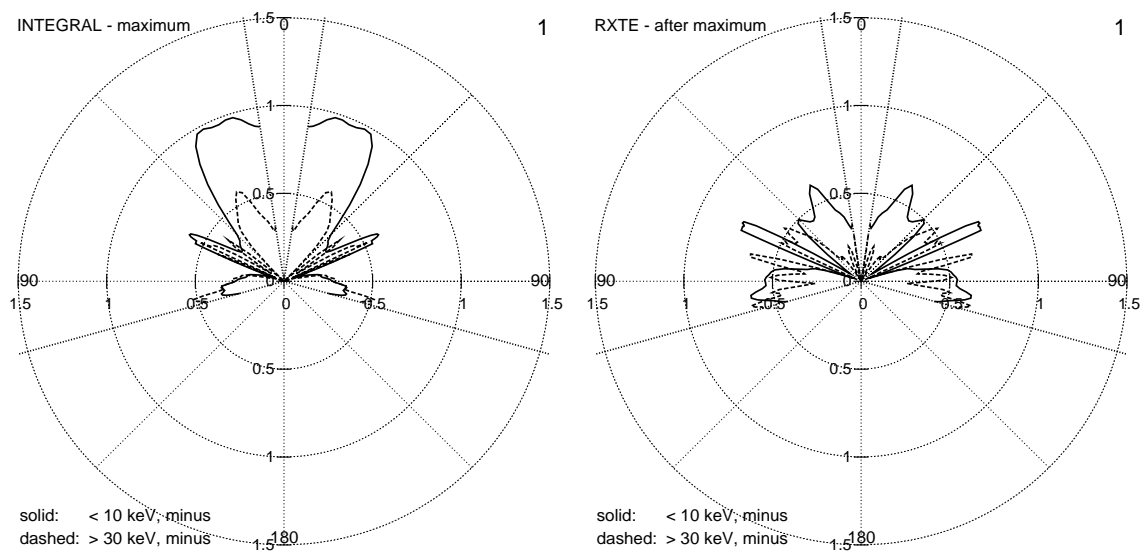

\centering
\includegraphics[width=0.4\textwidth,angle=270,clip=]{polar.A1.bw.eps_pages3}
\includegraphics[width=0.4\textwidth,angle=270,clip=]{polar.A1.bw.eps_pages4}
\caption{Polar diagrams of the intrinsic beam pattern for the {\tt minus} 
solution of solution 1 
for the same observations as shown in Fig.\,\ref{bptheta1}, for the softest 
and the hardest bands for an assumed $\Theta_{0}$ = 50\degr. Dotted lines 
are drawn to clearly show the range in which the emission is visible for
the observer. 
%The upper panel shows the beam patterns for the {\tt plus} 
%solution, the lower panel the {\tt minus} solution.
}
\label{bpintrA1}
\end{figure*}
%______________________________________________________________

%%______________________________________________ intrinsic beam patterns A2
%\begin{figure*}
%\centering
%\includegraphics[width=0.4\textwidth,angle=270,clip=]{polar.A2.bw.eps_pages1}
%\includegraphics[width=0.4\textwidth,angle=270,clip=]{polar.A2.bw.eps_pages2}\\[3mm]
%\includegraphics[width=0.4\textwidth,angle=270,clip=]{polar.A2.bw.eps_pages3}
%\includegraphics[width=0.4\textwidth,angle=270,clip=]{polar.A2.bw.eps_pages4}
%\caption{Polar diagrams of the intrinsic beam pattern for solution 2 
%for the same observations as shown in Figure \ref{bptheta2}, for the softest 
%and the hardest bands for an assumed $\Theta_{0}$ = 50\degr, with dotted lines
%showing the range in which the emission is visible. The beam patterns 
%from the two emission regions have been averaged in the range of overlap.
%The upper panel shows the beam patterns for the {\tt plus} 
%solution, the lower panel the {\tt minus} solution.
%}
%\label{bpintrA2}
%\end{figure*}
%______________________________________________________________

The observed asymptotic beam pattern differs from the intrinsic local beam 
pattern of the emission region because of relativistic light deflection 
near the neutron star. Making assumptions on the radius $r$ and mass 
$m$ of the neutron star, the asymptotic angle $\theta$ can be 
transformed into the intrinsic angle $\vartheta$ relative to the surface 
normal at the first magnetic pole to describe the local emission pattern. We 
use the canonical values $r = 10$~km and $m = 1.4~M_{\sun}$, 
thus the ratio between $r$ and the Schwarzschild radius $r_{\rm S}$ is 
$r/r_{\rm S}$ = 2.4. 
The intrinsic beam patterns of the {\tt minus} solution
are shown in Fig.\,\ref{bpintrA1} in polar diagrams.
% and \ref{bpintrA2}. 
%For solution 2, the two beam patterns in the overlap range
%are averaged by calculating the mean to obtain the total visible beam pattern
%shown in Figures \ref{bptheta2} and \ref{bpintrA2}.
As can be seen in these polar diagrams, the range for 
the visible angle is $\sim$10\degr\ -- 105\degr\ if $\Theta_{0}$ = 50\degr\ is 
assumed. 

\subsection{Possible origin of the emission}\label{emorigin}

In X-ray pulsars, a strong magnetic field funnels matter accreted from the
companion star onto the polar caps. The infalling particles deposit their 
energy in the atmosphere of the neutron star.
Heating due to accretion is balanced by radiative cooling through
bremsstrahlung emission, Compton scattering, and the so-called
cyclotron radiation after collisional excitation of electrons.  
In low-luminosity states, the accreted matter is decelerated in the 
atmosphere of the neutron star by Coulomb scattering. Emission can be seen
as pencil beams from the hot spots at polar caps \citep{1981A&A...102...97W}.
In high-luminosity states, however, the accretion rate is higher, and an
accretion column is believed to form. Plasma is decelerated by radiation 
pressure in the column, and a radiative shock forms above the neutron star 
surface \citep{1973Natur.246....1D,1976MNRAS.175..395B}.
Above the radiative shock, plasma is in free fall, while below the 
discontinuity, there is a region of nearly stagnant plasma from which photons 
escape from the sides of the column in a fan beam.

In addition, a luminous halo might form around the accretion column 
\citep{1973ApJ...179..585D,1988SvAL...14..390L} and
radiation from the polar cap and the halo can also be scattered in the upper 
accretion stream \citep{1985A&A...144..485S,1991ApJ...369..179B}.
As the accretion stream is delimited by magnetic field lines of the neutron 
star, it opens up wide far from the neutron star. Therefore, the emission from 
the upper part of the accretion stream can dominate the observed flux and can 
also screen the polar caps and the halo. Due to relativistic light deflection, 
emission from one pole can be deflected to the antipodal direction 
\citep{1983ApJ...274..846P,1988ApJ...325..207R,1995MNRAS.277.1177L,2001ApJ...563..289K}.
All these effects can modify the local beam pattern and thus have an effect on
the pulse profile of the X-ray pulsar.
\citet{1981A&A...102...97W} 
have modeled the pulse profiles of X-ray pulsars by assuming hot spots and fan 
beams, although \exo\ is not included in their list of selected X-ray pulsars. 
They show that, in general, sharp minima can be seen in the 
pulse profiles at the moment when the fan beam rotates behind the horizon of 
the neutron star and when it reappears on the side of the neutron star facing 
the observer. This is exactly what seems to be happening in \exo\ at phases 
0.1 and 0.6 as the decomposition method has shown.

Kraus et al.\ (\citeyear{2003ApJ...590..424K}; in prep.)
have modeled beam patterns and pulse profiles for
medium-luminosity X-ray pulsars assuming an accretion column with 
energy-dependent local beaming of radiation, a luminous halo formed by 
illumination of the neutron star surface, and magnetic scattering
in the upper accretion stream. 
They show that the emission from the upper accretion stream can
dominate the local beam pattern for $\theta >$ 120\degr. Halo emission has 
a maximum at $\theta \approx$ 30\degr\ -- 60\degr, with significant contribution to 
the beam pattern for photon energies below $\sim$5~keV. If one assumes
isotropic emission from the accretion column below the discontinuity, 
the halo emission becomes comparable or negligible relative to the column 
emission at $\ga$10~keV. However, if the radiation below the shock is
beamed downwards, the halo remains more luminous than the accretion column
and the upper stream up to $\sim$30~keV.
The decomposition has shown that the beam patterns show dominant 
soft emission for $\theta <$ 60\degr, which can be interpreted as halo 
emission, while the emission $\theta >$ 150\degr\ most likely arises from 
the upper accretion stream (see Fig.\,\ref{bptheta1}, lower panel).
%, \ref{bptheta2}). 

\section{Summary}\label{summary}

We performed pulse-profile decomposition with data of \exo\ taken
during the giant outburst of 2006 by \rxte\ and \integral. This is the fourth 
source after Cen~X-3, Her~X-1, and A~0535+26 to which this method has been 
applied. 
Each of the asymmetric pulse profiles of \exo\ at different luminosities in 
various energy bands are decomposed in two symmetric pulse profiles that account
for emission from the two emission regions of the system. 

We find that the magnetic field of the neutron star is moderately 
distorted. The observer sees a part of the emission from each of the two 
emission regions, but these parts do not overlap. 
%In solution 2, on the contrary,
%most of the emission from both emission regions is visible. This 
%solution suggests an extreme geometry for the neutron star, in which the
%two magnetic poles are located close to the equator of the neutron star 
%spanning an angle of about 90\degr\ between each other. The beam patterns of 
%the two emission regions do not seem to be perfectly equal, causing small
%modulations of the pulse profiles.
%
We suggest that the main peak at phase 0.2 in the observed pulse profiles 
can be attributed to harder emission from the upper accretion stream
of the second pole, which can be observed while the pole is on the other side 
of the neutron star because of relativistic light bending around the neutron star
(`anti-pencil'). 
However, the main peak also has a considerable contribution 
from the halo emission of the first pole, which is closer to the line of sight 
than the second. Between phases $\sim$0.95 and $\sim$0.1, both poles are located 
on the rear side of the neutron star. The sharp minima seen at phases 0.0 and 0.6
are caused when the first pole is about to re- and disappear at the horizon of the 
neutron star, and its accretion column is seen from the side. 

Our analysis has disentangled the emission components of the neutron star,
which in total lead to the observed asymmetric, energy-, and 
luminosity-dependent pulse profiles of \exo. 
It will allow us to perform detailed analyses 
of, e.g., the pulse phase-resolved spectra with reliable interpretation of the 
differences in the spectral parameters, hence shedding light on the physical 
processes in the system responsible for the observed emission.

\begin{acknowledgements}
This research is based on observations with INTEGRAL, an ESA project with 
instruments and science data centre funded by ESA member states (especially 
the PI countries: Denmark, France, Germany, Italy, Switzerland, Spain), Poland, 
and with the participation of Russia and the USA.
This research has made use of data obtained from the High Energy Astrophysics 
Science Archive Research Center (HEASARC), provided by NASA's Goddard Space 
Flight Center.
This work was supported by DFG grant SA 1777 1/1.
\end{acknowledgements}

\bibliographystyle{aa} % style aa.bst
\bibliography{../../bibtex/xb,../../bibtex/xpuls,../../bibtex/exo2030,../../bibtex/xraytel}

\begin{thebibliography}{42}
\expandafter\ifx\csname natexlab\endcsname\relax\def\natexlab#1{#1}\fi

\bibitem[{{Angelini} {et~al.}(1989){Angelini}, {Stella}, \&
  {Parmar}}]{1989ApJ...346..906A}
{Angelini}, L., {Stella}, L., \& {Parmar}, A.~N. 1989, \apj, 346, 906

\bibitem[{{Basko} \& {Sunyaev}(1976)}]{1976MNRAS.175..395B}
{Basko}, M.~M. \& {Sunyaev}, R.~A. 1976, \mnras, 175, 395

\bibitem[{{Bildsten} {et~al.}(1997){Bildsten}, {Chakrabarty}, {Chiu}, {Finger},
  {Koh}, {Nelson}, {Prince}, {Rubin}, {Scott}, {Stollberg}, {Vaughan},
  {Wilson}, \& {Wilson}}]{1997ApJS..113..367B}
{Bildsten}, L., {Chakrabarty}, D., {Chiu}, J., {et~al.} 1997, \apjs, 113, 367

\bibitem[{{Blum} \& {Kraus}(2000)}]{2000ApJ...529..968B}
{Blum}, S. \& {Kraus}, U. 2000, \apj, 529, 968

\bibitem[{{Brainerd} \& {Meszaros}(1991)}]{1991ApJ...369..179B}
{Brainerd}, J.~J. \& {Meszaros}, P. 1991, \apj, 369, 179

\bibitem[{{Bulik} {et~al.}(1995){Bulik}, {Riffert}, {Meszaros}, {Makishima},
  {Mihara}, \& {Thomas}}]{1995ApJ...444..405B}
{Bulik}, T., {Riffert}, H., {Meszaros}, P., {et~al.} 1995, \apj, 444, 405

\bibitem[{{Caballero} {et~al.}(2010){Caballero}, {Kraus}, {Santangelo},
  {Sasaki}, \& {Kretschmar}}]{Caballero2010}
{Caballero}, I., {Kraus}, U., {Santangelo}, A., {Sasaki}, M., \& {Kretschmar},
  P. 2010, \aap, submitted

\bibitem[{{Coe} {et~al.}(1988){Coe}, {Payne}, {Longmore}, \&
  {Hanson}}]{1988MNRAS.232..865C}
{Coe}, M.~J., {Payne}, B.~J., {Longmore}, A., \& {Hanson}, C.~G. 1988, \mnras,
  232, 865

\bibitem[{{Corbet} \& {Levine}(2006)}]{2006ATel..843....1C}
{Corbet}, R.~H.~D. \& {Levine}, A.~M. 2006, The Astronomer's Telegram, 843, 1

\bibitem[{{Davidson}(1973)}]{1973Natur.246....1D}
{Davidson}, K. 1973, \nat, 246, 1

\bibitem[{{Davidson} \& {Ostriker}(1973)}]{1973ApJ...179..585D}
{Davidson}, K. \& {Ostriker}, J.~P. 1973, \apj, 179, 585

\bibitem[{{Frontera} \& {Dalfiume}(1989)}]{1989ESASP.296...57F}
{Frontera}, F. \& {Dalfiume}, D. 1989, in ESA Special Publication, Vol. 296,
  Two Topics in X-Ray Astronomy, Volume 1: X Ray Binaries. Volume 2: AGN and
  the X Ray Background, ed. {J.~Hunt \& B.~Battrick}, 57--69

\bibitem[{{Gehrels} {et~al.}(2004){Gehrels}, {Chincarini}, {Giommi}, {Mason},
  {Nousek}, {Wells}, {White}, {Barthelmy}, {Burrows}, {Cominsky}, {Hurley},
  {Marshall}, {M{\'e}sz{\'a}ros}, {Roming}, {Angelini}, {Barbier}, {Belloni},
  {Campana}, {Caraveo}, {Chester}, {Citterio}, {Cline}, {Cropper}, {Cummings},
  {Dean}, {Feigelson}, {Fenimore}, {Frail}, {Fruchter}, {Garmire}, {Gendreau},
  {Ghisellini}, {Greiner}, {Hill}, {Hunsberger}, {Krimm}, {Kulkarni}, {Kumar},
  {Lebrun}, {Lloyd-Ronning}, {Markwardt}, {Mattson}, {Mushotzky}, {Norris},
  {Osborne}, {Paczynski}, {Palmer}, {Park}, {Parsons}, {Paul}, {Rees},
  {Reynolds}, {Rhoads}, {Sasseen}, {Schaefer}, {Short}, {Smale}, {Smith},
  {Stella}, {Tagliaferri}, {Takahashi}, {Tashiro}, {Townsley}, {Tueller},
  {Turner}, {Vietri}, {Voges}, {Ward}, {Willingale}, {Zerbi}, \&
  {Zhang}}]{2004ApJ...611.1005G}
{Gehrels}, N., {Chincarini}, G., {Giommi}, P., {et~al.} 2004, \apj, 611, 1005

\bibitem[{{Jahoda} {et~al.}(1996){Jahoda}, {Swank}, {Giles}, {Stark},
  {Strohmayer}, {Zhang}, \& {Morgan}}]{1996SPIE.2808...59J}
{Jahoda}, K., {Swank}, J.~H., {Giles}, A.~B., {et~al.} 1996, in Society of
  Photo-Optical Instrumentation Engineers (SPIE) Conference Series, Vol. 2808,
  Society of Photo-Optical Instrumentation Engineers (SPIE) Conference Series,
  ed. {O.~H.~Siegmund \& M.~A.~Gummin}, 59--70

\bibitem[{{Janot-Pacheco} {et~al.}(1988){Janot-Pacheco}, {Motch}, \&
  {Pakull}}]{1988A&A...202...81J}
{Janot-Pacheco}, E., {Motch}, C., \& {Pakull}, M.~W. 1988, \aap, 202, 81

\bibitem[{{Klochkov} {et~al.}(2007){Klochkov}, {Horns}, {Santangelo},
  {Staubert}, {Segreto}, {Ferrigno}, {Kretschmar}, {Kreykenbohm}, {La Barbera},
  {Masetti}, {McCollough}, {Pottschmidt}, {Sch{\"o}nherr}, \&
  {Wilms}}]{2007A&A...464L..45K}
{Klochkov}, D., {Horns}, D., {Santangelo}, A., {et~al.} 2007, \aap, 464, L45

\bibitem[{{Klochkov} {et~al.}(2008){Klochkov}, {Santangelo}, {Staubert}, \&
  {Ferrigno}}]{2008A&A...491..833K}
{Klochkov}, D., {Santangelo}, A., {Staubert}, R., \& {Ferrigno}, C. 2008, \aap,
  491, 833

\bibitem[{{Kraus}(2001)}]{2001ApJ...563..289K}
{Kraus}, U. 2001, \apj, 563, 289

\bibitem[{{Kraus} {et~al.}(1996){Kraus}, {Blum}, {Schulte}, {Ruder}, \&
  {Meszaros}}]{1996ApJ...467..794K}
{Kraus}, U., {Blum}, S., {Schulte}, J., {Ruder}, H., \& {Meszaros}, P. 1996,
  \apj, 467, 794

\bibitem[{{Kraus} {et~al.}(1995){Kraus}, {Nollert}, {Ruder}, \&
  {Riffert}}]{1995ApJ...450..763K}
{Kraus}, U., {Nollert}, H.-P., {Ruder}, H., \& {Riffert}, H. 1995, \apj, 450,
  763

\bibitem[{{Kraus} {et~al.}(2003){Kraus}, {Zahn}, {Weth}, \&
  {Ruder}}]{2003ApJ...590..424K}
{Kraus}, U., {Zahn}, C., {Weth}, C., \& {Ruder}, H. 2003, \apj, 590, 424

\bibitem[{{Krimm} {et~al.}(2006){Krimm}, {Barthelmy}, {Gehrels}, {Markwardt},
  {Palmer}, {Sanwal}, \& {Tueller}}]{2006ATel..861....1K}
{Krimm}, H., {Barthelmy}, S., {Gehrels}, N., {et~al.} 2006, The Astronomer's
  Telegram, 861, 1

\bibitem[{{Leahy}(1991)}]{1991MNRAS.251..203L}
{Leahy}, D.~A. 1991, \mnras, 251, 203

\bibitem[{{Leahy} \& {Li}(1995)}]{1995MNRAS.277.1177L}
{Leahy}, D.~A. \& {Li}, L. 1995, \mnras, 277, 1177

\bibitem[{{Lund} {et~al.}(2003){Lund}, {Budtz-J{\o}rgensen}, {Westergaard},
  {Brandt}, {Rasmussen}, {Hornstrup}, {Oxborrow}, {Chenevez}, {Jensen},
  {Laursen}, {Andersen}, {Mogensen}, {Rasmussen}, {Om{\o}}, {Pedersen},
  {Polny}, {Andersson}, {Andersson}, {K{\"a}m{\"a}r{\"a}inen}, {Vilhu},
  {Huovelin}, {Maisala}, {Morawski}, {Juchnikowski}, {Costa}, {Feroci},
  {Rubini}, {Rapisarda}, {Morelli}, {Carassiti}, {Frontera}, {Pelliciari},
  {Loffredo}, {Mart{\'{\i}}nez N{\'u}{\~n}ez}, {Reglero}, {Velasco}, {Larsson},
  {Svensson}, {Zdziarski}, {Castro-Tirado}, {Attina}, {Goria}, {Giulianelli},
  {Cordero}, {Rezazad}, {Schmidt}, {Carli}, {Gomez}, {Jensen}, {Sarri},
  {Tiemon}, {Orr}, {Much}, {Kretschmar}, \& {Schnopper}}]{2003A&A...411L.231L}
{Lund}, N., {Budtz-J{\o}rgensen}, C., {Westergaard}, N.~J., {et~al.} 2003,
  \aap, 411, L231

\bibitem[{{Lyubarskii} \& {Syunyaev}(1988)}]{1988SvAL...14..390L}
{Lyubarskii}, Y.~E. \& {Syunyaev}, R.~A. 1988, Soviet Astronomy Letters, 14,
  390

\bibitem[{{McCollough} {et~al.}(2006){McCollough}, {Turler}, {Willis}, \&
  {Shaw}}]{2006ATel..868....1M}
{McCollough}, M.~L., {Turler}, M., {Willis}, D., \& {Shaw}, S.~E. 2006, The
  Astronomer's Telegram, 868, 1

\bibitem[{{Motch} \& {Janot-Pacheco}(1987)}]{1987A&A...182L..55M}
{Motch}, C. \& {Janot-Pacheco}, E. 1987, \aap, 182, L55

\bibitem[{{Parmar} {et~al.}(1989{\natexlab{a}}){Parmar}, {White}, \&
  {Stella}}]{1989ApJ...338..373P}
{Parmar}, A.~N., {White}, N.~E., \& {Stella}, L. 1989{\natexlab{a}}, \apj, 338,
  373

\bibitem[{{Parmar} {et~al.}(1989{\natexlab{b}}){Parmar}, {White}, {Stella},
  {Izzo}, \& {Ferri}}]{1989ApJ...338..359P}
{Parmar}, A.~N., {White}, N.~E., {Stella}, L., {Izzo}, C., \& {Ferri}, P.
  1989{\natexlab{b}}, \apj, 338, 359

\bibitem[{{Pechenick} {et~al.}(1983){Pechenick}, {Ftaclas}, \&
  {Cohen}}]{1983ApJ...274..846P}
{Pechenick}, K.~R., {Ftaclas}, C., \& {Cohen}, J.~M. 1983, \apj, 274, 846

\bibitem[{{Reynolds} {et~al.}(1993){Reynolds}, {Parmar}, \&
  {White}}]{1993ApJ...414..302R}
{Reynolds}, A.~P., {Parmar}, A.~N., \& {White}, N.~E. 1993, \apj, 414, 302

\bibitem[{{Riffert} \& {Meszaros}(1988)}]{1988ApJ...325..207R}
{Riffert}, H. \& {Meszaros}, P. 1988, \apj, 325, 207

\bibitem[{{Riffert} {et~al.}(1993){Riffert}, {Nollert}, {Kraus}, \&
  {Ruder}}]{1993ApJ...406..185R}
{Riffert}, H., {Nollert}, H.-P., {Kraus}, U., \& {Ruder}, H. 1993, \apj, 406,
  185

\bibitem[{{Soffel} {et~al.}(1985){Soffel}, {Herold}, {Ruder}, \&
  {Ventura}}]{1985A&A...144..485S}
{Soffel}, M., {Herold}, H., {Ruder}, H., \& {Ventura}, J. 1985, \aap, 144, 485

\bibitem[{{Ubertini} {et~al.}(2003){Ubertini}, {Lebrun}, {Di Cocco}, {Bazzano},
  {Bird}, {Broenstad}, {Goldwurm}, {La Rosa}, {Labanti}, {Laurent}, {Mirabel},
  {Quadrini}, {Ramsey}, {Reglero}, {Sabau}, {Sacco}, {Staubert}, {Vigroux},
  {Weisskopf}, \& {Zdziarski}}]{2003A&A...411L.131U}
{Ubertini}, P., {Lebrun}, F., {Di Cocco}, G., {et~al.} 2003, \aap, 411, L131

\bibitem[{{Wang} \& {Welter}(1981)}]{1981A&A...102...97W}
{Wang}, Y. \& {Welter}, G.~L. 1981, \aap, 102, 97

\bibitem[{{White} {et~al.}(1983){White}, {Swank}, \&
  {Holt}}]{1983ApJ...270..711W}
{White}, N.~E., {Swank}, J.~H., \& {Holt}, S.~S. 1983, \apj, 270, 711

\bibitem[{{Wilson} {et~al.}(2005){Wilson}, {Fabregat}, \&
  {Coburn}}]{2005ApJ...620L..99W}
{Wilson}, C.~A., {Fabregat}, J., \& {Coburn}, W. 2005, \apjl, 620, L99

\bibitem[{{Wilson} {et~al.}(2008){Wilson}, {Finger}, \&
  {Camero-Arranz}}]{2008ApJ...678.1263W}
{Wilson}, C.~A., {Finger}, M.~H., \& {Camero-Arranz}, A. 2008, \apj, 678, 1263

\bibitem[{{Wilson} {et~al.}(2002){Wilson}, {Finger}, {Coe}, {Laycock}, \&
  {Fabregat}}]{2002ApJ...570..287W}
{Wilson}, C.~A., {Finger}, M.~H., {Coe}, M.~J., {Laycock}, S., \& {Fabregat},
  J. 2002, \apj, 570, 287

\bibitem[{{Winkler} {et~al.}(2003){Winkler}, {Courvoisier}, {Di Cocco},
  {Gehrels}, {Gim{\'e}nez}, {Grebenev}, {Hermsen}, {Mas-Hesse}, {Lebrun},
  {Lund}, {Palumbo}, {Paul}, {Roques}, {Schnopper}, {Sch{\"o}nfelder},
  {Sunyaev}, {Teegarden}, {Ubertini}, {Vedrenne}, \&
  {Dean}}]{2003A&A...411L...1W}
{Winkler}, C., {Courvoisier}, T.~J.-L., {Di Cocco}, G., {et~al.} 2003, \aap,
  411, L1

\end{thebibliography}

%\begin{thebibliography}{}
%   \bibitem[year]{name}lalala
%\end{thebibliography}

\appendix

\section{The decomposition method}

The pulse-profile decomposition method has been developed and first 
presented by \citet{1995ApJ...450..763K}. Here we quickly summarize 
the method and the steps in the analysis.

\subsection{Decomposition into two single-pole pulse profiles}
\label{meth_decomp}

\begin{figure*}
\centering
\includegraphics[width=0.455\textwidth,clip=]{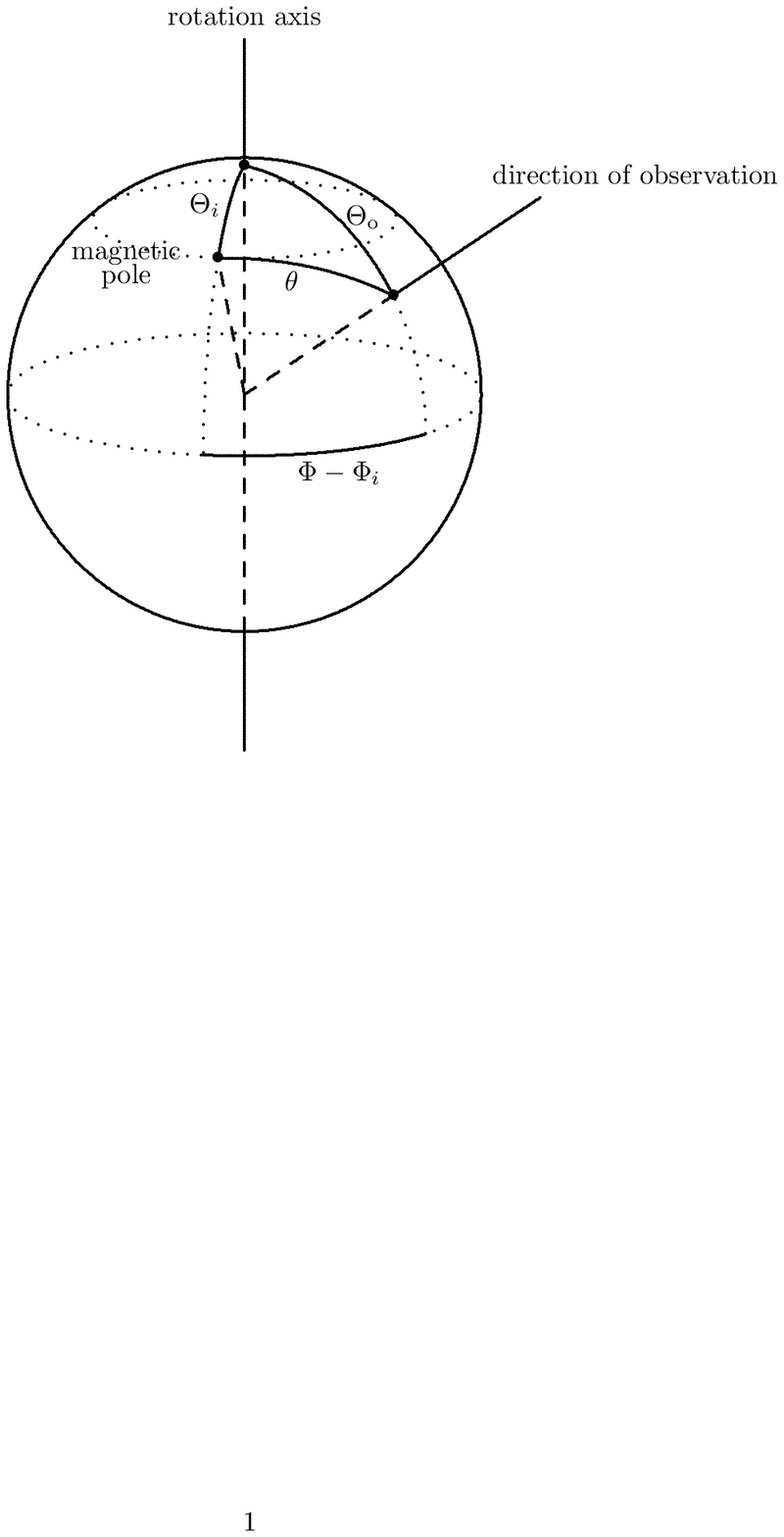}
\includegraphics[width=0.4\textwidth,clip=]{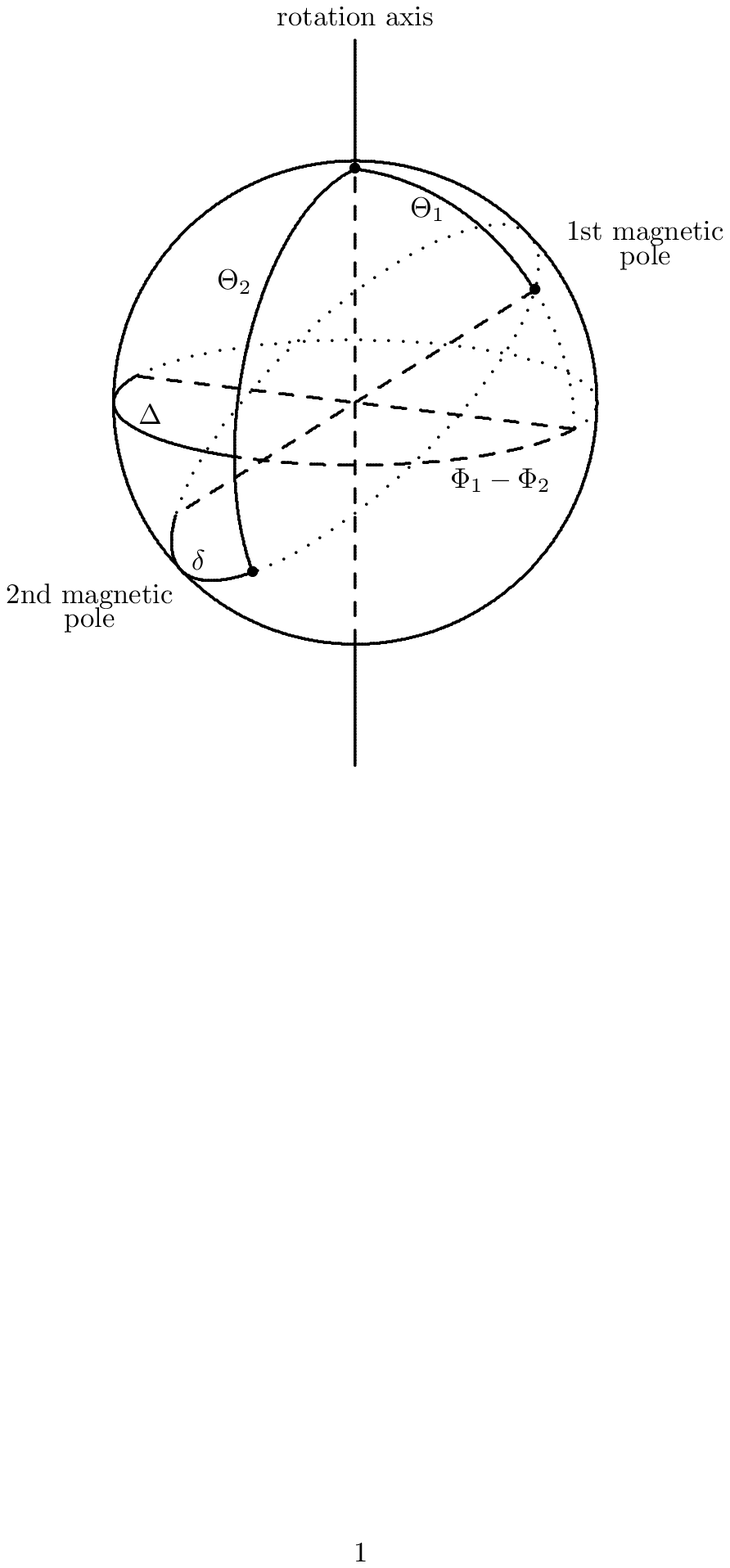}
\caption{Intrinsic geometry of the neutron star. Figures taken from
\citet{1995ApJ...450..763K}}
\label{angles}
\end{figure*}

Let $\theta$ be the angle between the direction of the line of sight and
the axis through one magnetic pole. 
Figure \ref{angles} shows the configuration of the neutron star, showing
the positions of the magnetic poles with respect to the rotation axis. 
The polar angles of the two magnetic poles are called $\Theta_{1}$ and 
$\Theta_{2}$. 
As the neutron star rotates, $\theta$ changes with rotation angle $\Phi$.
For the emission of each pole, there are two symmetry points at $\Phi_{i}$
and $\Phi_{i} + \pi\ (i = 1, 2)$. 
For each pole, the relation between $\theta$, $\Theta_{i}$, and $\Phi_{i}$
can be obtained using the spherical triangle in Fig.\,\ref{angles} (left):
\begin{equation}\label{theta}
\cos\theta = \cos\Theta_{0}\cos\Theta_{i} + \sin\Theta_{0}\sin\Theta_{i}\cos(\Phi-\Phi_{i}). 
\end{equation}
We assume that there is an offset $\delta$ between the
second magnetic pole and the antipodal position of the first magnetic pole,
therefore, there will be a phase offset for the rotation angles $\Phi_{i}$
from a symmetric configuration which can be defined as 
\begin{equation}\label{delta}
\Delta = \pi - (\Phi_{1} - \Phi_{2}). 
\end{equation}
For an ideal dipole field
$\Theta_{1} + \Theta_{2} = \pi$, $\Delta = 0$, and $\delta = 0$.
With the polar angles $\Theta_{1}$, $\Theta_{2}$, and the offset $\Delta$ in 
the rotation angles, we have a complete set of parameters to describe the 
geometry of the neutron star.   

To find the contributions from each magnetic pole, we perform Fourier analysis
of the observed total pulse profiles. We model it as a sum of two symmetric 
functions $f_{1}$ and $f_{2}$ and search for the values for their symmetry
points $\Phi_{1}$ and $\Phi_{2}$, respectively. 
In principle we are able to find a set of $f_{1}$ and $f_{2}$ for any chosen
$\Phi_{1}$ and $\Phi_{2}$. 
However, since we deal with functions that describe astronomically
observed emission, the following criteria need to be fulfilled:
\begin{enumerate}
\item {\it Positive flux:} The symmetric functions $f_{1}$ and $f_{2}$ must
not have negative values because they model the flux of an astronomical object.
\item {\it No ripples:} The functions $f_{1}$ and $f_{2}$
should show no small-scale features that cancel out in the sum. As the two
functions correspond to pulse profiles of single-poles that emit independently,
they ought not to have features that match exactly. Also, the single-pole
pulse profiles are supposedly not more complicated than the total pulse 
profile.
\item {\it Same geometry:} We have pulse profiles from one source in 
different energy bands and, ideally, from more than one observation. 
The observed pulse profiles are energy and 
luminosity dependent in most cases. Since the emission arises from only one 
object, the symmetry points must be the same for the decompositions 
of all available data.
\end{enumerate}
The actual search for reasonable decompositions is performed using the 
parameters $\Phi_{1}$ and $\Delta$. The second symmetry point $\Phi_{2}$ can
then be obtained from Eq.\,\ref{delta}. The decompositions of the pulse 
profiles of different energy bands and different observations will not yield 
identical values for $\Phi_{1}$ and $\Delta$. 
Therefore we have to define interesting regions in the parameter space of 
$\Phi_{1}$ and $\Delta$ and look at the single-pole pulse profile for similar
values of $\Phi_{1}$ and $\Delta$ for each energy band and observation to 
decide if they can be declared as one consistent set of emission. 
This means that the single-pole pulse profiles have to have similar values of
$\Phi_{1}$ and $\Delta$ and ought to be similar from one energy band to 
the other in one observation. Once one has found such a set of decomposition 
for all energy bands and observations
for similar $\Phi_{1}$ and $\Delta$, one can compute the corresponding beam
patterns for each pole.

\subsection{Reconstruction of the beam pattern}\label{meth_bp}

Two beam patterns are obtained from single-pole pulse profiles as 
functions of the rotation angles 
$\Phi$ (for one single-pole pulse profile) and $\tilde{\Phi}$ 
(for the other single-pole pulse profile).
During one revolution of the neutron star, the angle $\theta$ between the 
first magnetic pole and the line of sight changes with the phase. 
There may be an interval during one phase, i.e., a range of the angle $\theta$, 
in which the observer sees emission from both poles.
The same is true for the second magnetic pole. The range covered by the angle 
$\theta$ is in general different for each pole.
There may, however, be a certain subrange of values of $\theta$
that occur for each of the poles. Then,
at some phase $\Phi$, the observer looks onto the first pole at angle
$\theta$ and at some different phase $\tilde{\Phi}$, the observer looks onto
the second pole at the same angle $\theta$.
A simple example would be two antipodal poles that pass through the line
of sight ($\theta$ = 0) at $\Phi$ = 0 and at $\tilde{\Phi}$ = 0.5, respectively.       
If, in addition, the two emission regions of the neutron star have the
same beam pattern, the emission seen at $\theta$ from the first pole
(phase $\Phi$) is the same as the emission seen at $\theta$ from the second
pole (phase $\tilde{\Phi}$). 
This means that the visible beam patterns of 
the two poles must have identical parts at different pulse phases.
If we find such parts of the beam patterns of the two single-pole 
pulse profiles we can overlay them and get a relation between 
$\cos(\Phi-\Phi_{1})$ and $\cos(\tilde{\Phi}-\Phi_{2})$. From this relation, 
we can derive the positions of the magnetic poles $\Theta_{1}$ and 
$\Theta_{2}$ as functions of the direction of the observation $\Theta_{0}$. 
The angle $\Theta_{0}$ needs to be determined independently 
in other studies of the source.

%----------------------------------------------------------- angles
%______________________________________________________________

%____________________________________________ asymptotic beam patterns (phase)
\begin{figure*}
\centering
\hspace{8mm}{\footnotesize\sf Solution 1}\\[-3mm]
\includegraphics[width=0.5\textwidth,angle=270,clip=]{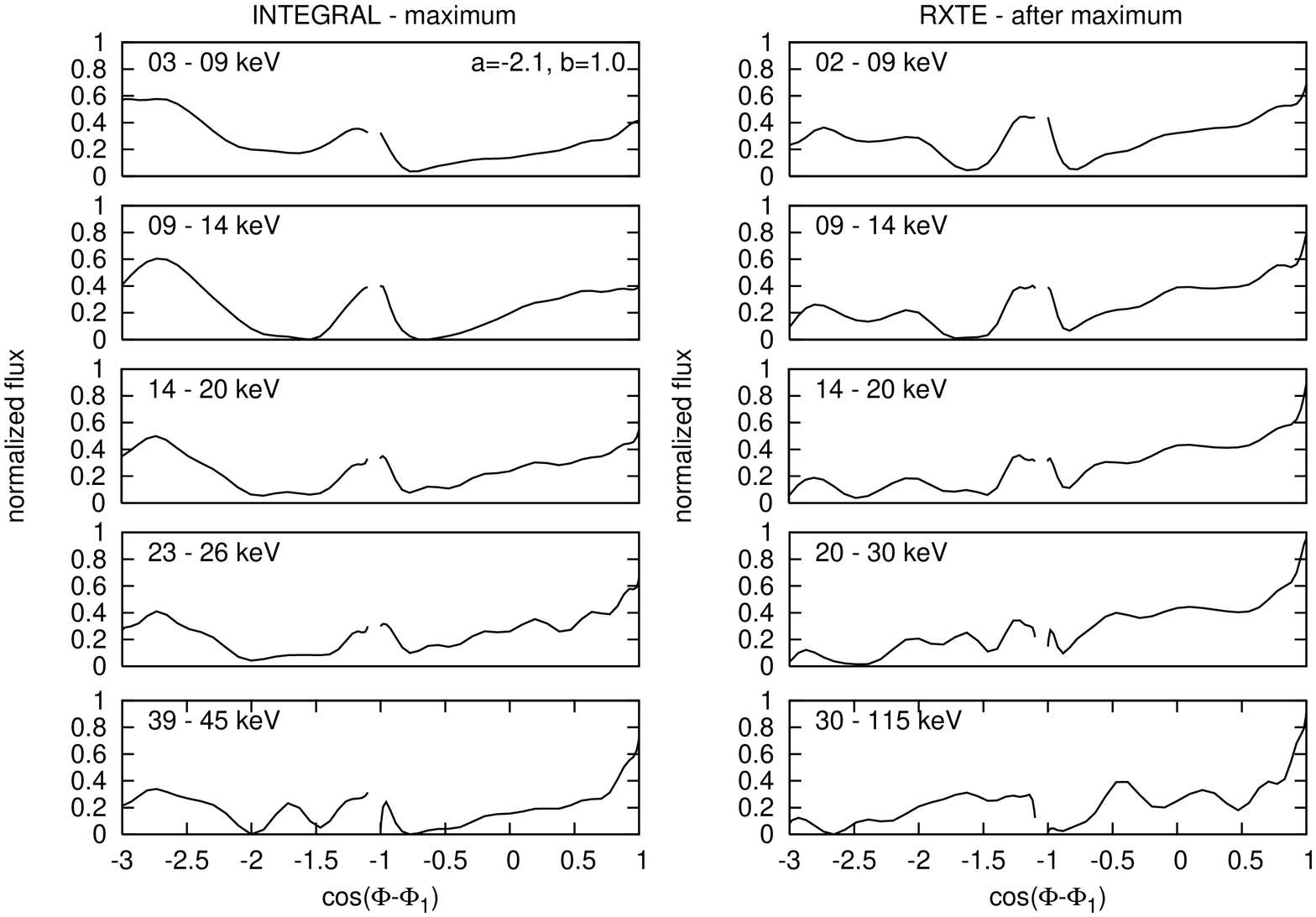}
\\[1mm]
\hspace{8mm}{\footnotesize\sf Solution 2}\\[-3mm]
\includegraphics[width=0.5\textwidth,angle=270,clip=]{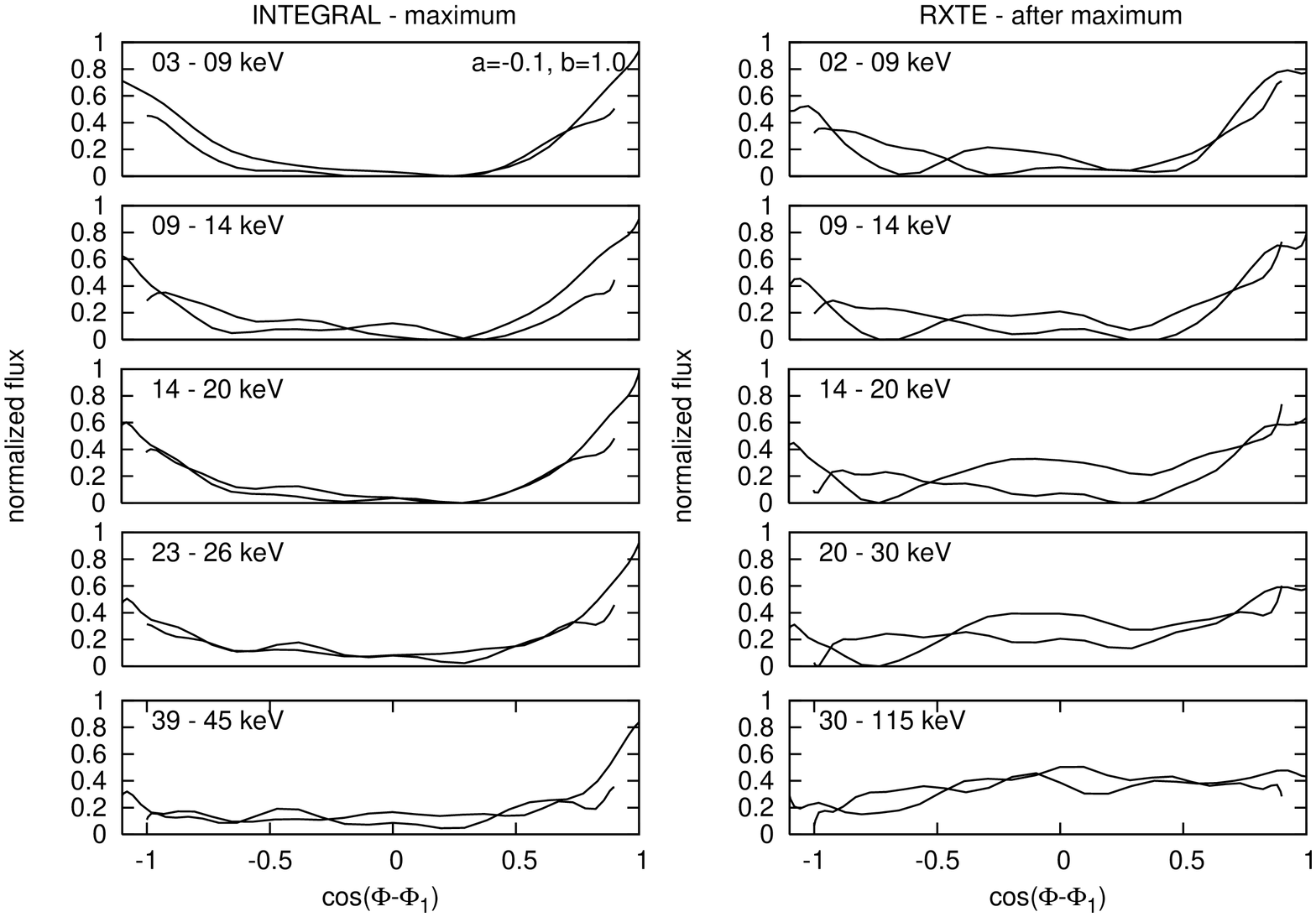}
\caption{Beam patterns from the two emission regions seen by the observer 
plotted over $\cos(\Phi-\Phi_{1})$. The left panel shows the beam patterns of 
solution 1 of the observation near maximum as shown in Figure \ref{sppbp}.
The right panel shows a selection of additional data from the observation 
after maximum with similar energy bands. 
For solution 1 (upper panel), 
In this case no overlap of the beam patterns for the two emission regions 
is found. 
For solution 2 
(lower panel), the beam patterns seem to overlap over a wide range, but show 
small differences. 
For the parameters $a, b$, see Section \ref{asymbp}.}
\label{bpobs}
\end{figure*}
%______________________________________________________________

%_________________________ pulse profile of A2 assuming same emission pattern
\begin{figure*}
\centering
\includegraphics[width=0.209\textwidth,angle=270,clip=]{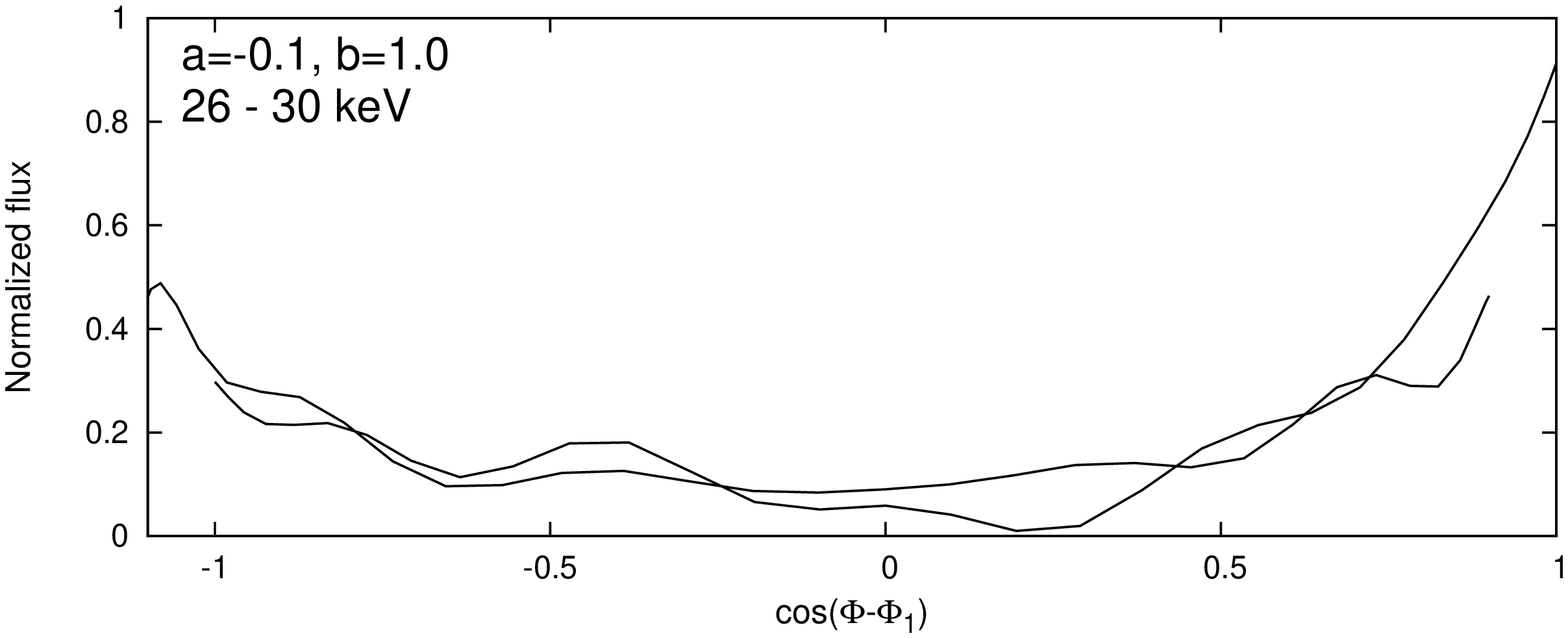}
\includegraphics[width=0.209\textwidth,angle=270,clip=]{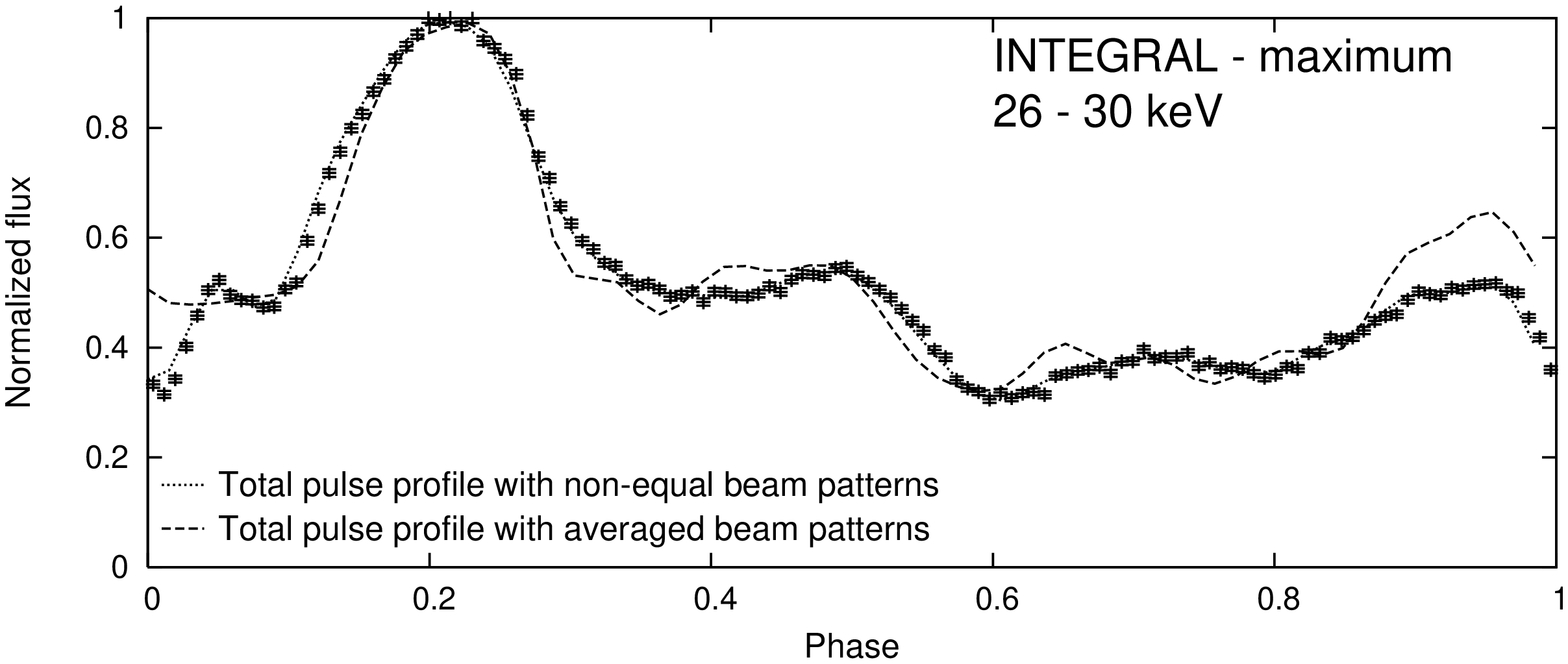}
\caption{
{\it Left:}
Beam patterns from the two emission regions plotted over 
$\cos(\Phi-\Phi_{1})$ for the energy band 26 -- 30~keV of the \integral\ 
observation near maximum for solution 2. 
For this solution, the beam patterns 
seem to overlap over a wide range, but show small differences between each
other. 
{\it Right:}
Observed total pulse profile (data points with errors) with 
total pulse profiles reconstructed from the not perfectly matching beam 
patterns of solution 2 for the two emission regions (dotted) and averaged beam 
patterns assuming equal local emission pattern for the two regions (dashed).
}
\label{totalppA2}
\end{figure*}
%______________________________________________________________

\end{document}